\documentclass[aps,prd,12pt,a4paper,superscriptaddress,preprintnumbers,nofootinbib,floatfix]{revtex4-1}
\usepackage[top=2.9cm, bottom=2.1cm, left=2cm, right=2cm]{geometry}
\usepackage[english]{babel}
\usepackage{amsmath,amssymb,amsfonts, dsfont}
\usepackage{subfigure}
\usepackage{graphicx}
\usepackage{color}
\usepackage{hyperref}
\usepackage{dcolumn}% Align table columns on decimal point
\usepackage{bm}% bold math
\usepackage{mathtools}
\usepackage{amsthm}
\usepackage{bbm}
\usepackage{pxfonts}
\usepackage[vcentermath]{youngtab}
\usepackage[all]{xy}
\usepackage{accents}
\usepackage{cases}
\usepackage{comment}
\usepackage{placeins}
\usepackage{xspace}
\usepackage{cancel} % to make the barred text notation
\usepackage{slashed}
\usepackage{soul}
\usepackage{feynmp}
\usepackage{epstopdf}
\usepackage{xcolor}

\definecolor{orange}{cmyk}{0,0.5,1,0}
\definecolor{rossoCP3}{cmyk}{0,.88,.77,.40}
\definecolor{graa}{rgb}{0.8,0.8,0.8}
\definecolor{blaa}{rgb}{0.2,0.2,0.6}
\hypersetup{
	colorlinks, 
	bookmarksopen, 
	bookmarksnumbered,
	citecolor=blaa, 		%color of links to bibliography
	linkcolor=rossoCP3,	%color of internal links
	urlcolor=rossoCP3,			%color of external links 
}
\newcommand{\be}{\begin{equation}}
\newcommand{\ee}{\end{equation}}

\newcommand{\SU}{\mathrm{SU}} 
\newcommand{\Sp}{\mathrm{Sp}}

\newcommand{\parenbar}[2][4]{%
  \mkern#1mu
  \sbox0{$#2$}%
  \makebox[0pt][r]{\raisebox{\ht0}{$\scriptscriptstyle($}}%
  \overline{\mkern-#1mu#2\mkern-1mu}%
  \makebox[0pt][l]{\raisebox{\ht0}{$\scriptscriptstyle)$}}%
  \mkern1mu
}

\begin{document}

%%%%%%%%%%%%%%TITLE AFFILIATIONS ETC%%%%%%%%%%%%%%%%%%%%%%%%%%%%%%%%%%%%%%%%%%%%%%%%%%%%%%%%%%%%
 
\title{\texorpdfstring{\Large\color{rossoCP3}   Baryogenesis via Elementary Goldstone Higgs Relaxation}{}}
 
\author{Helene {\sc Gertov}}
\email{gertov@cp3.sdu.dk} 
\affiliation{{\color{rossoCP3} {CP}$^{ \bf 3}${-Origins}} \& the Danish Institute for Advanced Study {\color{rossoCP3}\rm{Danish IAS}},  University of Southern Denmark, Campusvej 55, DK-5230 Odense M, Denmark.}

\author{Lauren {\sc Pearce}}
\email{lpearce@umn.edu} 
\affiliation{Department of Physics and Astronomy, Valparaiso University, Valparaiso, IN 46383 USA}
\affiliation{William I. Fine Theoretical Physics Institute, School of Physics and Astronomy, University of Minnesota, Minneapolis, MN 55455 USA}

\author{Francesco {\sc Sannino}}
\email{sannino@cp3.dias.sdu.dk}
\affiliation{{\color{rossoCP3} {CP}$^{ \bf 3}${-Origins}} \& the Danish Institute for Advanced Study {\color{rossoCP3}\rm{Danish IAS}},  University of Southern Denmark, Campusvej 55, DK-5230 Odense M, Denmark.}

\author{Louis {\sc Yang}}
\email{louis.yang@physics.ucla.edu}  
\affiliation{Department of Physics and Astronomy, University of California, Los
Angeles, California 90095-1547, USA}

%%%%%%%%%%%%%%%%%%%%%%%%%%%%%%%%%%%%%%%%%%%%%%%%%%%%%%%%%%%%%%%%%%%%%%%%%%

\begin{abstract}
We extend the relaxation mechanism  to the  Elementary Goldstone Higgs framework.  Besides studying  the allowed parameter space of the theory we add the minimal ingredients needed for the framework to be phenomenologically viable. 
The very nature of the extended Higgs sector allows to consider very flat scalar potential directions along which the relaxation mechanism can be implemented. This fact translates into wider regions of applicability of the relaxation mechanism when compared to the Standard Model Higgs case. 

Our results show that, if the electroweak scale is not fundamental but radiatively generated, it is possible to generate the observed matter-antimatter asymmetry via the relaxation mechanism.   \\[.3cm] 
{\footnotesize  \it Preprint: CP$^3$-Origins-2016-002 DNRF90 \& DIAS-2016-2 FTPI-MINN-16/02}
\end{abstract}
\maketitle
\newpage
%\tableofcontents
%\newpage

%%%%%%%%%%%%%%%%%%%%%%%%
%%%%%%%%%%%%%%%%%%%%%%%%

\section{Introduction}

% Edits by Francesco Sannino, 17/01/2016
% Edits by Helene Gertov, 24/01/2016

The discovery of the Higgs boson crowns the Standard Model (SM) of particle interactions as one of the most successful description of physical phenomena below or at around the electroweak (EW) scale. However, several puzzles remain unexplained such as the nature of dark matter, neutrino masses and mixing as well as the cosmological matter-antimatter asymmetry of the universe.  Solutions to any of these puzzles  generically requires introduction of new physics beyond the SM. 

Here we focus our attention on the important cosmological mystery of how the observable universe came to be dominated by an excess of matter over antimatter.  The necessary conditions for baryogenesis are well-known \cite{Sakharov:1967dj} and several models of baryogenesis exist (for a review, see e.g. \cite{Cline:2006ts}). Among these, an appealing scenario involves the relaxation of a scalar or pseudo scalar field in the post-inflationary universe.  Such fields can acquire large vacuum expectation values due to flat potentials \cite{Bunch:1978yq,*Linde:1982,*Starobinsky:1994bd} or by being trapped in a quasi-stable minimum.  After inflation, such fields relax to their equilibrium values via a coherent motion, and higher dimensional operators can couple the time-dependent condensate to baryon and/or lepton number \cite{Cohen:1987vi,*Dine:1990fj}.  This can be done with the Higgs field (\cite{Kusenko:2014lra,Pearce:2015nga,Yang:2015ida}) or an axion field (\cite{Kusenko:2014uta,*Adshead:2015jza}).    Similar models have been constructed using quintessence fields (\cite{Li:2001st,DeFelice:2002ir}) and MSSM flat directions (\cite{Chiba:2003vp,Takahashi:2003db,Takahashi:2003db,Kamada:2012ht}).  A novel feature of the Higgs scenario is that the chemical potential depends on the time-derivative of the VEV-squared, which as we discuss, resolves some difficulties with producing an asymmetry of the correct sign throughout the observable universe.  An additional advantage of this scenario is that it makes use of fields whose existence is either known or well-motivated.

However the SM Higgs sector is far from established and could hide new exciting physics. In fact, several alternative paradigms have been put forward that are not only as successful as the SM in reproducing the experimental results, but also can simultaneously address some of the remaining experimental puzzles. 

The Elementary Goldstone Higgs (EGH) paradigm established in \cite{Alanne:2014kea,Gertov:2015xma}   allows one to disentangle the vacuum expectation of the {\it elementary} Higgs sector from the EW scale \cite{Alanne:2014kea}.  Here the Higgs sector symmetry is larger than the minimally required symmetry needed to spontaneously break the EW gauge symmetry. Furthermore, the physical Higgs emerges as a pseudo Nambu Goldstone Boson (pNGB). A welcome feature is that once the SM gauge and fermion sectors are embedded in the larger symmetry, one discovers that calculable radiative corrections automatically induce the proper breaking of the EW symmetry by naturally aligning the vacuum in the pNGB Higgs direction.  In this way the EW scale is {not fundamental}  but radiatively induced\footnote{The EGH setup is profoundly different from the composite (Goldstone) Higgs scenario \cite{Kaplan:1983fs, *Kaplan:1983sm, *Cacciapaglia:2014uja}. 
The main differences being: i)  the elementary case is amenable to perturbation theory; ii) it is straightforward to endow the SM fermions with mass terms; iii) it is possible to immediately consider Grand Unified Theory extensions \cite{Alanne:2015fqh}.}.  The template  Higgs sector leading to the  $\mathrm{SU}(4) \to \mathrm{Sp}(4)$ pattern of chiral symmetry breaking  was first introduced in \cite{Appelquist:1999dq,*Duan:2000dy,*Ryttov:2008xe}.

In this work, we successfully marry the EGH paradigm and the relaxation leptogenesis  scenario \cite{Kusenko:2014lra,Yang:2015ida}.  There are at least two motivations for this marriage: first, we observe that in a model such as the EGH scanario, which has an extended scalar sector, there are naturally new scalar-field directions along which one can implement the relaxation mechanism; secondly, the relative freedom in the overall potential flatness translates into a wider region of applicability of the approximations and effective theory used to derive successful baryogenesis (as compared to the SM Higgs case).

The structure of this paper is as follows: we begin by reviewing the EGH model, as introduced in \cite{Alanne:2014kea, Gertov:2015xma}, with particular emphasis on the scalar sector and the Yukawa sector.  We then review the Higgs relaxation scenario, focusing on the modifications necessary due to the extended Higgs sector.  Finally, we present an analysis of the available parameter space in which Higgs relaxation leptogenesis occurs when marrying it to the EGH paradigm.

Although we use a specific template to perform our analysis, the general results and features are expected to hold for generic realisations of successful baryogenesis via EGH driven leptogensis scenarios.

\section{Elementary Goldstone Higgs: A brief review}

The EGH scenario \cite{Alanne:2014kea, Gertov:2015xma} necessarily extends the SM Higgs sector symmetry. A working template uses a linear realisation with  $\SU(4)$ symmetry breaking spontaneously  to $\Sp(4)$.  The SM Higgs doublet is now part of the $\SU(4)/\Sp(4)$ coset, while the EW symmetry,  $\SU(2)_L\times$U$(1)_Y$, is embedded in $\SU(4)$.  

The relaxation leptogenesis mechanism \cite{Kusenko:2014lra,Pearce:2015nga,Yang:2015ida}  uses the scalar sector of the theory at very high energies. We therefore start by reviewing the scalar sector of the theory.  
 
 The SM Higgs boson is identified with one of the Goldstone bosons which  acquires mass via a slight vacuum misalignment mechanism induced by quantum corrections. The misalignment is due mostly to top-induced quantum corrections  \cite{Alanne:2014kea,Gertov:2015xma}  and therefore we will neglect here  the EW gauge sector corrections.  
 
The most general vacuum structure of the theory can be parametrised by an angle $0\leq\theta\leq\pi/2$  \cite{Alanne:2014kea}
 and
   \be E_\theta =\cos\theta \,E_B +  \sin\theta\, E_H= -E^T_\theta\,,\label{eq:E}\ee
where the two independent  vacuum directions $E_B$ and $E_H$ are 
\be  E_B =
\begin{pmatrix}
i\sigma_2 & 0\\
0 & -i \sigma_2
\end{pmatrix}, \quad  E_H=
\begin{pmatrix}
0 & 1\\
-1& 0
\end{pmatrix}\,.\label{vacua}
\ee
 The alignment angle $\theta$ is determined by radiative corrections after having constrained the model to reproduce the experimental results. It was found in  \cite{Gertov:2015xma}  that the model naturally prefers small values of $\theta$, privileging a pNGB nature of the Higgs. 
 
We note that because $\mathrm{SU}(4)$ is broken radiatively through corrections from top and gauge interactions, the strength of SU(4) breaking increases at higher scales.  Therefore, there is no high scale in which it is appropriate to neglect SU(4) breaking.

%%%%%%%%%%%%%%%%%%%%%%%%%%%%%%%%%
%%%%%%%%%%%%%%%%%%%%%%%%%%%%%%%%%
\subsection{Scalar and Fermionic sector}\label{sec:scalars}
%%%%%%%%%%%%%%%%%%%%%%%%%%%%%%%%%
%%%%%%%%%%%%%%%%%%%%%%%%%%%%%%%%%
In the minimal scenario described  above, the  scalar sector is codified by   \be \begin{split}
M   %& = e^{\frac{i}{f} 2 \sqrt{2} \sum_i \left(\Pi_i(x) + i \tilde \Pi_i(x)\right) X_i}  \frac{1}{2}\left(\sigma+i  \Theta \right)\, E\\
     & =\left[\frac{1}{2} \left( \sigma + i\, \Theta\right) + \sqrt{2}\, ( \Pi_i+i \,\tilde \Pi_i) \,X^i_\theta \right] E_\theta\,,  \\ \end{split}
\label{eq:higgssector}
\ee
where  $X^i_\theta$  ($i = 1, \ldots, 5$)  are the broken generators associated to the breaking of  $\mathrm{SU}(4)$ to $\mathrm{Sp}(4)$, reported 
in Appendix A of \cite{Gertov:2015xma}, and the five $\Pi_i$ are the five Goldtone bosons of the theory, where after symmetry breaking, the first three become the longitudinal  components  of the gauge bosons, the fourth is the observed Higgs, and the last is a dark matter candidate. 
The full $\mathrm{SU}(4)$ invariant (tree-level) scalar potential can be found in  \cite{Gertov:2015xma}. %

Having introduced the scalar sector of the model, we turn our attention to the fermionic sector.  Our focus here is two-fold: first, we have explained above how the top-sector is responsible for setting $\theta$, and therefore indirectly affects the vacuum and the scalar sector more generally, and secondly, the leptogenesis scenario produces the asymmetry  through the excess production of neutrinos, which involves electroweak interactions between fermions.

We construct the Yukawa sector of the  theory by introducing EW gauge invariant operators that explicitly break  the $\mathrm{SU}(4)$ global symmetry and
correctly  reproduce the SM fermion masses and mixing. 
First, we formally accommodate each one of the SM fermion families in the fundamental representation of $\mathrm{SU}(4)$, namely
\be  \mathbf{L}_{\alpha}=\begin{pmatrix} L, &  \tilde \nu, & \tilde \ell
\end{pmatrix}_{\alpha L}^T\sim \mathbf{4}, \qquad  
\mathbf{Q}_{i}=\begin{pmatrix} Q, &  \tilde q^u, & \tilde q^d
\end{pmatrix}_{i \,L}^T \sim \mathbf{4},\label{eq:ql}
\ee
where $\alpha=e,\mu,\tau$ and $i=1,  2, 3$ are generation indices  and the tilde indicates the charge conjugate fields of the Right Handed (RH) fermions, that is, for instance, 
$\tilde\nu_{\alpha L}\equiv (\nu_{\alpha R})^c$, $\tilde \ell_{\alpha L}\equiv (\ell_{\alpha R})^c$, $L_{\alpha L}\equiv (\nu_{\alpha L}, \ell_{\alpha L})^T$ and similarly for the quark fields. Notice that a RH neutrino  $\nu_{\alpha R}$ for each family must be introduced in order to define $\mathbf{L}_{\alpha}$ which transforms according to the fundamental irrepresentation of $\mathrm{SU}(4)$.

Given the embedding of quarks and leptons in $\mathrm{SU}(4)$, we now construct a Yukawa mass term for the SM fermions. 
 For this we make use of $\mathrm{SU}(4)$ spurion fields \cite{Galloway:2010bp} $P_a$ and $\overline{P}_a$, where  $a=1,2$ is an $\mathrm{SU}(2)_L$ index. They transform   as  $\parenbar[1]{P}_a \rightarrow (u^\dagger)^T \,\parenbar[1]{P}_a \,u^\dagger$, with $u \in \mathrm{SU}(4)$. We have
\be \begin{split}
 P_1= 
\frac{1}{\sqrt{2}}
\begin{pmatrix}
\bf{0}_2 & \tau_3\\
-\tau_3 & \bf{0}_2
\end{pmatrix}\,, 
& 
\quad  P_2= 
\frac{1}{\sqrt{2}}
\begin{pmatrix}
\bf{0}_2 & \tau^-\\
-\tau^+ & \bf{0}_2
\end{pmatrix}\,,
\end{split}\ee
\be \overline P_1=\frac{1}{\sqrt{2}}
\begin{pmatrix}
\bf{0}_2 & \tau^+\\
-\tau^- & \bf{0}_2
\end{pmatrix}
\,,
\quad \overline  P_2= 
\frac{1}{\sqrt{2}}
\begin{pmatrix}
\bf{0}_2 & \overline \tau_3\\
- \overline \tau_3\ & \bf{0}_2
\end{pmatrix}\,, \\
\ee
with
\be \tau^\pm=\frac{\sigma_1\pm i\, \sigma_2}{2}, \,\, \tau_3=\frac{\bf{1}_2+\sigma_3}{2}, \quad\mbox{and} \quad \overline \tau_3=\frac{\bf{1}_2- \sigma_3}{2}\,. \ee

Then, using $P_{1,2}$ and $\overline P_{1,2}$, we may write Yukawa couplings for the SM fermions which preserve the $\mathrm{SU}(2)_L$ gauge symmetry:
%%%%
 \begin{eqnarray}
-\mathcal{L}^\text{Yukawa} &= & \frac{Y^u_{i j}}{\sqrt{2}} \,\left(\mathbf{Q}^T_{i} \, P_a \, \mathbf{Q}_{j} \right)^\dagger Tr\left[P_a \, M\right] + 
 \frac{Y^d_{i j}}{\sqrt{2}}\, \left(\mathbf{Q}^T_{i}  \,\overline P_a\,  \mathbf{Q}_{j} \right)^\dagger Tr\left[ \overline P_a\, M\right]\nonumber\\ 
         &+& \frac{Y^\nu_{\alpha\beta}}{\sqrt{2}} \,\left(\mathbf{L}^T_{\alpha} \, P_a \, \mathbf{L}_{\beta} \right)^\dagger Tr\left[P_a \, M\right] + 
 \frac{Y^\ell_{\alpha\beta}}{\sqrt{2}}\, \left(\mathbf{L}^T_{\alpha}  \,\overline P_a\,  \mathbf{L}_{\beta} \right)^\dagger Tr\left[ \overline P_a\, M\right]\,+\,\text{h.c.}\label{LYuk1}
\end{eqnarray}
with the Yukawa matrices of quarks and leptons chosen in agreement with  experimental measurements.
\mathversion{normal}
This Lagrangian explicitly breaks the $\mathrm{SU}(4)$ global symmetry mentioned above, and therefore, it also contributes fixing the parameter $\theta$ which interpolates between the otherwise equivalent vacuum structures.  In fact, in terms of the SM quark and lepton fields, Eq.~(\ref{LYuk1}) can be written as
 \begin{eqnarray}
-\mathcal{L}^\text{Yukawa} &= & Y^u_{i j} \left( Q_{i L} \,  \tilde q^u_{j L} \right)^\dagger_a\ Tr\left[P_a \, M\right] + 
  Y^d_{i j} \, \left( Q_{i L} \,  \tilde q^d_{j L} \right)^\dagger_a Tr\left[ \overline P_a\, M\right]\nonumber\\ 
         &+& Y^\nu_{\alpha\beta} \,\left(L_{\alpha L} \,  \tilde \nu_{\beta L} \right)^\dagger_a Tr\left[P_a \, M\right] + 
  Y^\ell_{\alpha\beta}\, \left(L_{\alpha L} \,  \tilde \ell_{\beta L}  \right)^\dagger_a Tr\left[ \overline P_a\, M\right]\,+\,\text{h.c.}\label{LYuk2}
\end{eqnarray}
where 
\begin{align}
	Tr\left[P_{1}M\right]&=\frac{-1}{\sqrt{2}}\left(\sigma\sin\theta+\Pi_{4}\cos\theta+i\Theta\sin\theta-i\tilde{\Pi}_{4}\cos\theta+i\Pi_{3}+\tilde{\Pi}_{3}\right),\label{eq:neutrino_Y}\\
	Tr\left[P_{2}M\right]&=\frac{1}{\sqrt{2}} \left( i\Pi_1 +\Pi_2+\tilde\Pi_1-i\tilde\Pi_2  \right).
\end{align}
Therefore, after EW symmetry breaking, the SM fermions acquire the masses
\be m_F = y_F \frac{f \sin \theta}{\sqrt{2}}\,, \ee
with $f= \langle \sigma \rangle$ at low energies and $y_F$ being the SM Yukawa coupling of quarks and leptons in the fermion mass basis. 
Comparing this expression with the corresponding SM prediction $m_{F,SM}=\frac{y_F}{\sqrt{2}}v_{ew}$
we see that $f$ and $\theta$ must satisfy the phenomenological constraint
\be f\sin \theta\;=\;v_{\rm EW}\;\simeq\; 246 \mbox{ GeV} \,.\label{eq:ew}\ee
Notice that a Dirac mass for neutrinos is generated as well. Ref.~\cite{Gertov:2015xma} also investigated the parameter space at low energy and found that (when keeping the masses of the scalars below five TeV) the most frequent value for $f$ is $\overline f\;=\;13.9^{+2.9}_{-2.1}\mbox{ TeV}$ corresponding to $\overline\theta\;=\;0.018^{+0.004}_{-0.003}$. Although these are the most common values that give the appropriate electroweak phenomenology, the points of parameter space which satisfy the electroweak constraints vary significantly in values for $f$ and $\sin(\theta)$.  This is because there are quite a few couplings in the $\mathrm{SU}(4)$ potential.  To generate an acceptable electroweak phenomenology at values of $f$ and $\sin(\theta)$ significantly different than these, it is likely necessary to fine-tune at least some of the parameters in the $\mathrm{SU}(4)$ potential.

This model does not naturally generate a Majorana mass term for the RH neutrino fields; however, one can be explicitly added.  This provides an explicit breaking of the $\mathrm{SU}(4)$ symmetry, but preserves the
EW gauge group and gives the standard seesaw mechanism.  Although this is not strictly speaking necessary for the EGH boson model, we include it in our analysis here.  This is because a successful leptogenesis model must involve lepton-number violating terms, and following \cite{Kusenko:2014lra,Pearce:2015nga,Yang:2015ida}, we will make use of the neutrino-sector Majorana mass term.

In this case, the most general mass Lagrangian for the leptons is
\be -\mathcal{L}^{\text{lep}} =Y^\ell_{\alpha\beta}\, \frac{f\sin\theta}{\sqrt{2}} \,\,\overline \ell_{\alpha L}  \ell_{\beta R}+Y^\nu_{\alpha j} \frac{f\sin\theta}{\sqrt{2}}  \,\,\overline \nu_{\alpha L}  \nu_{j R}+ \frac{1}{2}\,(M_R)_{j k} \,  \overline \nu_{j R} \,(\nu_{k R})^c\,+\,\text{h.c.} 
\label{LYuk3}\ee
where $M_R$ is the Majorana mass term of the three RH neutrinos. 
% We require  $M_R$ to generate a small breaking of $\mathrm{SU}(4)$, that is
%\begin{equation}
%	\left| (M_R)_{j k}  \right|\;\ll\; f\,.
%\end{equation}
The couplings in Eq.~(\ref{LYuk3}) allow to generate at tree-level a Majorana mass term for the LH neutrinos, in a manner similar to the  standard type I seesaw extension of the SM \cite{Minkowski:1977sc,*Yanagida:1979as,*GellMann:1980vs,*Mohapatra:1979ia}.  This yields
\begin{equation}
 \mathcal{L}^\nu_{\text{mass}}\;=\;-\frac 1 2 \left( m_\nu \right)_{\alpha\beta}\,\overline\nu_{\alpha L}\,(\nu_{\beta L})^c\;+\;\text{h.c.}
\end{equation}
with
\begin{equation}
	m_\nu\;=\; - m_D\,M_R^{-1}\,m_D^T\quad \text{and}\quad m_D\;=\; Y^\nu\, \frac{f\sin\theta}{\sqrt{2}} \;=\; Y^\nu\, \frac{v_{\rm EW}}{\sqrt{2}}\,.
\end{equation}

One can hope that this Majorana mass term would be generated by embedding the EGH model into a larger model, perhaps a Grand Unified Theory.

%%%%%%%%%%%%%%%%%%%%%%%%%%%%%%
%%%%%%%%%%%%%%%%%%%%%%%%%%%%%%
\subsection{Radiative corrections}
%%%%%%%%%%%%%%%%%%%%%%%%%%%%%%
%%%%%%%%%%%%%%%%%%%%%%%%%%%%%%

\label{radiative}
Next we return our attention to the scalar sector.  The one-loop correction $\delta V(\Phi)$ to the scalar potential takes the general expression
	    \begin{equation}
		\label{eq:deltaV}
		\delta V(\Phi)=\frac{1}{64\pi^2}\mathrm{Str}\left[{\cal M}^4 (\Phi) \left(\log\frac{{\cal M}^2(\Phi)}
		    {\mu^2}-C\right)\right] %+V_{\mathrm{GB}},
	    \end{equation}
where in this case $\Phi\equiv(\sigma,\,\Pi_4)$ denotes  the background scalar fields that we expect to lead to the correct vacuum alignment of the theory
and  ${\cal M}(\Phi) $ is the corresponding  tree-level mass matrix. The  supertrace, $\mathrm{Str}$, is defined as
\begin{equation}
\mathrm{Str} = \sum_{\text{scalars}}-2\sum_{\text{fermions}}+3\sum_{\text{vectors}}.
\end{equation}
We have $C = 3/2$ for scalars and fermions and $C = 5/6$ for the gauge bosons, and $\mu_0$ is a reference renormalization scale.  
As explained above, the Yukawa sector terms explicitly break the global $\mathrm{SU}(4)$ symmetry and gauge interactions will also provide explicit symmetry breaking.
This explicit breaking will generate a nonzero mass term for the Goldstone bosons at the quantum  level and a mass mixing term between the $\Pi_4$ and the $\sigma$ fields. 

At very high energy scales the background-dependent  masses of all the scalars are the same, namely, $m \approx \sqrt{ \lambda }\sigma$, where $\lambda$ is a linear combination of several quartic couplings. The renormalisation scale is fixed as a constant at the energy scale of inflation. Taking  only the top and scalar corrections into account, the one-loop corrections to the potential  take the simple form
\begin{eqnarray}
	\delta V&=& \frac{\sigma^4}{64\pi^2} \left(7\lambda^2 \left( \log \frac{\lambda \sigma^2}{\mu^2}-\frac{2}{3} \right) -3 y_t^4 \sin^4\theta \left(\log\frac{y_t^2 \sin^2\theta\sigma^2}{2\mu^2} -\frac{3}{2} \right)\right).
\end{eqnarray}
in the direction of $\sigma$.   Thus the effective potential to one-loop order can be written as
\begin{eqnarray}
	V\left(\sigma\right)=\frac{\lambda_{\text{eff}}\left(\sigma\right)}{4}\sigma^{4}\label{eq:Veff}
\end{eqnarray}
where the effective quartic coupling is 
\begin{equation}
\lambda_{\text{eff}}\left(\sigma\right)=\lambda+\frac{4}{64\pi^{2}}\left[7\lambda^{2}\left(\log\left(\frac{\lambda\sigma^{2}}{\mu^{2}}\right)-\frac{3}{2}\right)-3y_{t}^{4}\sin^{4}\theta\left(\log\left(\frac{y_{t}^{2}\sin^{2}\theta\sigma^{2}}{2\mu^{2}}\right)-\frac{3}{2}\right)\right] \ .
\label{eq:lambda_eff}
\end{equation}
Here $y_t$ is the top quark Yukawa and $\lambda$ is not yet experimentally constrained. At lower energy, the one-loop potential is more involved and a detailed analysis of the one-loop potential at low energy can be found in \cite{Gertov:2015xma}. 
 In that paper the authors found that for $\theta$ around $0.018$ and $\lambda$ around $0.007$ there is a region of parameter space with the most “EW-favorable points”, but that it is by no means required for good EW behavior for theta and lambda to take these values.
In this work we will be primarily interested in the high energy regime since the scalar field will acquire a comparatively large vacuum expectation value.	Below, we will find that in order to produce a baryon asymmetry of the appropriate size, it is desirable to have a small coupling $\lambda_{\mathrm{eff}}$.  In order to ensure the stability of the potential at large $\sigma$, it may be necessary to tune both $\lambda$ and $\theta$ to be small, by choosing the parameters in the SU(4) potential appropriately. This is not a problem since, unlike in the SM, $\lambda_{\mathrm{eff}}$ is not set by the observed Higgs mass because of the enlarged scalar sector.

\subsection{The Physical Higgs}

For maximum clarity, we here pause to identify the physical Higgs boson states.  At low energy there is a mass mixing between the $\sigma$ and the $\Pi_4$ fields as mentioned earlier.  The mass eigenstates of this mixing are the two Higgs particles, $h$ and $H$, given by
\begin{eqnarray}
	\begin{pmatrix}
		\sigma\\
		\Pi_4
	\end{pmatrix}&=&
	\begin{pmatrix}
		\cos\alpha &-\sin\alpha\\
		\sin\alpha & \cos \alpha
	\end{pmatrix}
	\begin{pmatrix}
		h\\
		H
	\end{pmatrix}\,,\label{eq:Higgs}
\end{eqnarray}
where   $\alpha$ is the scalar mixing angle,  chosen in the interval $[0,\pi/2]$. 
The observed Higgs boson will be the lightest eigenstate with a mass 
\begin{eqnarray}
	 m_{h}=125.7\pm 0.4 \mbox{ GeV}\,\label{Hmass}
\end{eqnarray}  
which in \cite{Gertov:2015xma} was found that $\alpha$ is preferred to be very close to $\pi/2$; that is, the observed Higgs is mostly a pNGB.

As noted, though, at high energies these states are nearly degenerate in mass.  

\section{Relaxation-Leptogenesis framework}

Having introduced the EGH model, with particular attention to the scalar and fermionic sectors, we now introduce  the Higgs relaxation leptogenesis framework, which has been explored in the SM context in \cite{Kusenko:2014lra,Pearce:2015nga,Yang:2015ida}.  

We outline the important steps of relaxation leptogenesis as follows: First, we need a scalar (or pseudo-scalar) field with a large vacuum expectation value (VEV).  This can occur through quantum fluctuations during inflation, or the field may be trapped in a quasi-stable minimum.  Afterwards, the field relaxes to its equilibrium value.

During this relaxation, a chemical potential for lepton number may be induced via higher dimensional operators.  This lowers the energy of leptons and raises the energy of antileptons.  Lepton-number-violating processes, such as those mediated by the neutrino Majorana masses introduced above, produce an excess of leptons over antileptons.  These interactions can occur within the particle plasma produced during reheating \cite{Kusenko:2014lra,Yang:2015ida}, or during the decay of the Higgs condensate itself \cite{Pearce:2015nga}.  Here we focus on the first scenario as an illustrative example.

While we specifically consider the scalar sector here (which is of the most interest due to the extended scalar sector in the EGH model), we acknowledge that similar considerations apply to axion-like degrees of freedom, which have been explored in \cite{Kusenko:2014uta,*Adshead:2015jza}.

In our previous realisations of Higgs-relaxation leptogenesis, the relaxing field was identified with the SM Higgs, although we allowed for a modified potential at high scales.  The recent observation of the Higgs boson at the LHC sets the quartic coupling, although it is significantly modified at large scales (as described by the renormalization group equations)~\cite{Degrassi:2012ry}.  However, the EGH model has additional freedom as can be seen in Eq.~\eqref{eq:lambda_eff}.  We will show below that in order to generate the observed baryonic asymmetry, while remaining in the regime in which certain approximations are valid, the quartic coupling must be significantly smaller than the value preferred in the SM.  This is not phenomenologically problematic in the EGH scenario because the extended Higgs sector allows for additional flat directions.

\subsection{Large Initial Vacuum Expectation Value (VEV) of \texorpdfstring{$\sigma$}{Sigma}}

As we mentioned above, during inflation, scalar fields may acquire large vacuum expectation
values (VEVs) through quantum fluctuations: because relaxation via a coherent motion is a classical process, its time-scale may be significantly longer than those typical of quantum fluctuations (see \cite{Bunch:1978yq,*Linde:1982,*Starobinsky:1994bd}).  Concretely, quantum fluctations occur on a scale such that $V\left(\sigma_{I}\right)\sim H_{I}^{4}$ where $\sigma_{I}=\sqrt{\left\langle \sigma^{2}\right\rangle }$ is the scalar field vacuum expectation value and $H_{I}\equiv\sqrt{8\pi/3}\Lambda_{I}^{2}/M_{P}$ is the Hubble parameter during inflation.  The VEV rolls down classically to its minimum with the characteristic time scale 
\begin{equation}
\tau_{\text{roll}}\sim1/m_{\sigma,\text{eff}}\sim\left[\frac{d^{2}V\left(\sigma\right)}{d\sigma^{2}}\right]^{-1/2}.
\end{equation}
However, if $\tau_{\text{roll}}\gg H_{I}^{-1}$, there is insufficient
time between quantum fluctuations for the generated field VEV to roll down. In this case, the scalar field would develop a large VEV $\sim \sigma_I$ during inflation.  

An alternative scenario for starting with a large scalar vacuum expectation value is that the scalar field may be trapped in a quasi-stable minimum in the early universe; this is particularly well-motivated in scenarios in which the initial scalar VEVs are distributed stochastically (provided the scalar potential does, indeed, have a high-scale quasi-stable minimum).

The SM potential provides motivation for both scenarios: recent measurements of the Higgs mass suggest a rather flat potential, before turning over (and potentially becoming negative)~\cite{Degrassi:2012ry}.  The flat potential makes it easier for quantum fluctuations to generate a large VEV in the early universe; on the other hand, if the potential does turn over, higher-dimensional operators can stablize the potential in such a  way as to produce a quasi-stable minimum.

Here, though, we are interested in the EGH model, which has a different potential shape.  As noted, the physical Higgs boson $h$ and $H$ are mixtures of the $\sigma$ and $\Pi_4$ degrees of freedom, although there is an approximate rotational symmetry at high energies.  We will consider the case in which the field $\sigma$ acquires a large VEV during inflation within the effective potential given in Eq.~\eqref{eq:Veff}.  In fact at high energies $\sigma$ can be seen as simply the modulus of the scalar field and therefore it would not matter which direction one selects. Furthermore, as already explained, the effective quartic potential in this case is not fixed at low energies, because the mass of the pNGB Higgs emerges radiatively via top corrections and there are no sufficient experimental constraints yet to fix this overall coupling.  

Next, we address an issue which affects all relaxation leptogenesis models, which is discussed in more detail in \cite{Kusenko:2014lra,Yang:2015ida}.  Namely, in both ways of generating large scalar VEVs, different patches of the Universe generically have different values of $\sigma_{I}$ at the end of inflation. If the lepton asymmetry is linked to the initial VEV of $\sigma$, each patch of the Universe could have a different final asymmetry. This would result in unacceptably large baryonic isocurvature perturbations~\cite{Peebles1987,*1987ApJ...315L..73P,*Enqvist:1998pf,*Enqvist:1999hv,*Harigaya:2014tla}, which are constrained by the cosmic microwave background (CMB) observations~\cite{Ade:2013uln,*Ade:2015lrj}.

One solution to this problem, which was proposed in \cite{Kusenko:2014lra}, is to couple the Higgs sector to the inflaton in such a way as to suppress the growth of the VEV until the end of slow-roll inflation.  The resulting isocurvature perturbations are then on scales smaller than those which have been experimentally probed.  In the EGH model, we adapt this solution by coupling the $\sigma$ field to the inflaton $I$ via operators of the form
\begin{equation}
\mathcal{L}_{\sigma I}=c\frac{I^{n}}{M_{P}^{m+n-4}}\mathrm{Tr}\left[M^{+}M\right]^{m/2}.
\end{equation}
Such a non-renormalizable operator can be generated by integrating out heavy states in loops; we can envision that these states arise by heavy $\mathrm{SU}(4)$-preserving multiplets which arise when the EGH model is embedded into larger (perhaps grand unified) models.  In the early stages of inflation, the VEV of the inflaton $\left\langle I\right\rangle $
can be large (superplanckian) and gives a large effective mass $m_{\sigma,\text{eff}}\left(\left\langle I\right\rangle \right)$
to $\sigma$; this suppresses the quantum fluctuations of the $\sigma$ field. In
the later stages of inflation, $\left\langle I\right\rangle $ decreases
to a value such that $m_{\sigma,\text{eff}}\left(\left\langle I\right\rangle \right)\ll H_{I}$,
allowing a large VEV for $\sigma$ to develop. If the development of the
VEV occurs during the last $N_{\text{last}}$ $e$-folds of inflation,
the VEV reaches the average value 
\begin{equation}
\sigma_{0}=\min\left(\sigma_{I},\:\frac{H_{I}}{2\pi}\sqrt{N_{\text{last}}}\right).
\label{eq:initial_VEV}
\end{equation}
The resulting isocurvature perturbations appear only at the smallest
angular scales, and are not yet constrained for $N_{\text{last}}\lesssim8$.

While other solutions to this isocurvature problem were noted in \cite{Kusenko:2014lra}, we consider this one as an illustrative example which also allows the most freedom in parameter space.

\subsection{Relaxation of The \texorpdfstring{$\sigma$}{Sigma} Field}

When the inflation is over, the inflaton begins oscillating coherently as it decays; consequently, the universe behaves as if it is matter dominates.  During this epoch, the $\sigma$ field also relaxes from its starting value of $\sigma_{0}$ and oscillates around $\sigma=0$ (the minimum of Eq.~\eqref{eq:Veff}) with diminishing amplitude. The equation of motion for $\sigma\left(t\right)$ (where by an abuse of notation we use $\sigma(t)$ for the VEV of the $\sigma$ field) is 
\begin{equation}
\ddot{\sigma}+3H\left(t\right)\dot{\sigma}+\frac{dV\left(\sigma\right)}{d\sigma}=0.
\end{equation}
The Hubble parameter $H\left(t\right)$ is determined by the system
of differential equations
\begin{gather}
H\left(t\right)\equiv\frac{\dot{a}}{a}=\sqrt{\frac{8\pi}{3M_{P}^{2}}\left(\rho_{r}+\rho_{I}\right)},\\
\dot{\rho_{r}}+4H\left(t\right)\rho_{r}=\Gamma_{I}\rho_{I},
\end{gather}
where $\Gamma_{I}$ is the decay rate of inflaton, and $\rho_{I}=\Lambda_{I}^{4}e^{-\Gamma_{I}t}/a\left(t\right)^{3}$
and $\rho_{r}=\left(g_{*}\pi^{2}/30\right)T^{4}$ are the energy densities
of the inflaton field and the produced radiation respectively.  We ensure that the energy density of the $\sigma$ condensate never dominates the universe, so as to preserve that standard cosmological picture. Note that the maximum temperature during reheating and the reheat temperature can be estimated as  \cite{Weinberg:2008zzc} $T_\mathrm{max} \approx 0.618 \left(\Lambda_I^2 \, \Gamma_I \, m_{pl} / g_{*}\right)^{1/4}$ and \cite{Kolb:1990vq} $T_{R} \simeq \left(3/\pi^{3}g_{*}\right)^{1/4}\sqrt{\Gamma_{I}m_{pl}}$, respectively.

\subsection{Effective Chemical Potential}

We consider the following couplings between the lepton current and
the $\sigma$ field 
\begin{equation}
\mathcal{L}_{6}=-\frac{1}{M_{n}^{2}}Tr\left[M^{+}M\right]\partial_{\mu}j_{B}^{\mu}=-\frac{1}{M_{n}^{2}}\sigma^{2}\partial_{\mu}j_{B}^{\mu},
\label{eq:operator}
\end{equation}
where $M_{n}$ is a potentially new scale. This coupling does not break $\mathrm{SU}(4)$ symmetry, and is therefore consistent with the EGH picture.  As this is a higher-dimensional operator, it may be generated by integrating out heavy states; one obvious method is to expand the minimal EGH model introduced above with heavy RH states that couple to a gauge boson anomaly, as discussed in some detail in \cite{Yang:2015ida}.  Other possibilities for generating this operator, unique to EGH models, may exist.  We note that, to generate this effective operator, CP-violation is necessary~\cite{Shaposhnikov:1987tw,Shaposhnikov:1987pf}.

While this operator preserves $\mathrm{SU}(4)$, when the scalar VEV is evolving it breaks CPT; in fact, it is similar in form to the one considered in spontaneous baryogenesis scenarios~\cite{Cohen:1987vi,*Dine:1990fj}.  An integration by parts gives 
\begin{equation}
\mathcal{L}_{6}=\frac{1}{M_{n}^{2}}\left(\partial_{\mu}\sigma^{2}\right)j_{L}^{\mu}.
\end{equation}
For a patch of the Universe where the $\sigma$ field is approxiately spatially
homogeneous, the operator becomes
\begin{equation}
\mathcal{L}_{6}=\frac{1}{M_{n}^{2}}\left(\partial_{0}\sigma^{2}\right)j_{L}^{0}.
\end{equation}
When $\sigma$ is decreasing, this operator effectively raises the
energy of antiparticles, while lowering it for particles.  Within the equation of motion for the fermions, this term plays a role similar to that of an effective external chemical potential 
\begin{equation}
\mu_{0}=-\frac{\partial_{0}\sigma^{2}}{M_{n}^{2}}.
\end{equation}
In the presence of lepton-number-violating processes, the system will
favor the production of particles over antiparticles.

We emphasize that because the operator \eqref{eq:operator} depends on the VEV squared, a positive lepton number is produced everywhere as the Higgs field relaxes.  This is in contrast to many spontaneous baryogenesis models in which the effective chemical potential depends on $\partial_0 S$ (where $S$ is the scalar VEV), and in which the sign of the asymmetry depends on the initial sign of the VEV.  Consequently, in this model it is not necessary for the observable universe to be enclosed within a single patch of constant $\sigma$; instead we need only satisfy the isocurvature constraints mentioned above.  (We also note that the spatial variation in the field does not contribute to the charge asymmetry.)
\subsection{Lepton Number Violating Processes\label{sub:Lepton-Number-Violating}}

\begin{figure}
\centering{} \includegraphics{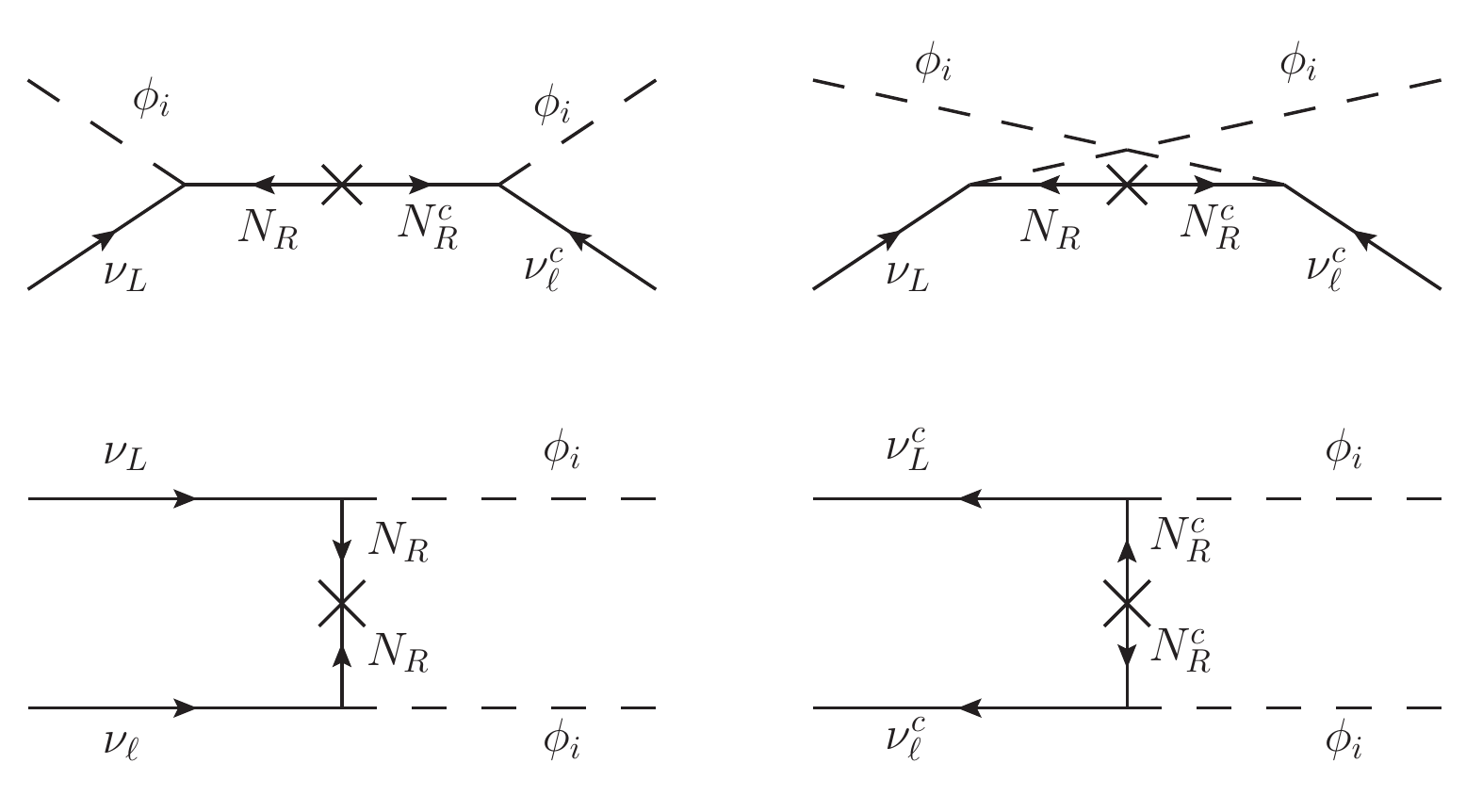} \protect\protect\caption{Some diagrams that contribute to lepton number violation via exchange
of a heavy Majorana neutrino, where $\phi_i$ can be $\Pi_{4}$, $\Theta$, $\tilde{\Pi}_{4}$, $\Pi_{3}$, $\tilde{\Pi}_{3}$, or $\sigma$, depending on which fields are in thermal equilibrium. 
\label{fig:lepton_violation}}
\end{figure}

This chemical potential alone will not yield any lepton asymmetry; successful leptogenesis additionally requires some lepton-number-violating process.  This was addressed above when we introduced Majorana masses in the neutrino sector. Therefore, we consider the standard seesaw mass matrix in Eq.~\eqref{LYuk3} for neutrinos (which requires a small breaking of $\mathrm{SU}(4)$, as discussed in~\cite{Gertov:2015xma}). The $L$ violating processes are (i) a left-handed neutrino converting into an anti-left-handed neutrino through exchange of a heavy Majorana neutrino; (ii) pair
production or annihilation of neutrinos or antineutrinos. These processes
are shown in the diagram in Fig.\ \ref{fig:lepton_violation}. 

With the introduction of the neutrino Majorana mass, several leptogenesis mechanisms become possible within the EGH model, beyond the Higgs relaxation leptogenesis considered here.  These include thermal leptogenesis~\cite{Fukugita:1986hr,Giudice:2003jh}, resonant leptogenesis~\cite{Pilaftsis:2003gt,Pilaftsis:2005rv}, and ARS-type leptogenesis~\cite{Akhmedov:1998qx}.  We consider a regime of parameter space in which these leptogenesis mechanisms are insufficient to generate the observed baryon asymmetry, but the asymmetry produced through the Higgs relaxation mechanism can account for the observed asymmetry.  In particular, we consider large Majorana mass $M_R$; in the Higgs relaxation mechanism, the smallness of the Majorana mass can be counteracted by a large $\partial_t \sigma^2$.  We also do not arrange the neutrino sector parameters so as to have resonant production of neutrinos.

The thermally averaged cross section of these processes $\left\langle \sigma v\right\rangle _{0}$ can be found in \cite{Yang:2015ida}.  This calculation can be easily extended to the EGH case, which will be discussed in Appendix \ref{app:16}. Due to the additional scalar fields that participate in these processes, the cross section is enhanced by a factor of 16 if all scalar fields are in thermal equilibrium,
\begin{align}
\left\langle \sigma v\right\rangle _{0} & =16\left[\left\langle \sigma\left(\nu_{L}\phi\leftrightarrow\bar{\nu}_{L}\phi\right)v\right\rangle _{0}+\left\langle \sigma\left(\nu_{L}\nu_{L}\leftrightarrow\phi\phi\right)v\right\rangle _{0}\right],\label{eq:sigma_R}
\end{align}
where $\phi$ stands for a scalar field that couples to $\nu_{L}$
with the Yukawa coupling $Y^{\nu}$.  We emphasize that all processes which violate lepton number contribute to producing the asymmetry in the presence of the nonzero chemical potential.  If heavier, weakly-interacting fields such as the $\sigma$ and the $\Theta$ are not in thermal equilibrium, then the cross section is scaled by $\left[ 1 - \sin^2\left(\theta\right)/2 \right]^2 \approx 1$.

% However, the heavier scalar particles will typically not be in thermal equilibrium; in this case, the cross section is dominated by the $\tilde{\Pi}_3$ and $\tilde{\Pi}_4$ contributions and includes a factor of $\cos(\theta)^2 \sim 1$.

For the $\nu_{L}\phi\leftrightarrow\bar{\nu}_{L}\phi$
process, the $s$-channel process has a resonance at $E\sim M_{R}$, which is generally ineffective because, in order to suppress standard leptogenesis, we ensure that $M_R$ is well above the energy scales probed by the VEV (and also above the reheat temperature). The center-of-mass cross section is 
\begin{multline}
\sigma_{CM}\left(\nu_{L}\phi\leftrightarrow\bar{\nu}_{L}\phi\right)=\dfrac{1}{16\pi}\dfrac{\left|Y^{\nu}\right|^{4}}{4}\left(M_{R}^{2}+\dfrac{\Gamma_{R}}{4}\right)\int_{-s}^{0}dt\,(s+t)\\
\times\left[\dfrac{1}{A^{2}+C^{2}}+\dfrac{1}{B^{2}+C^{2}}+\dfrac{2(AB+C^{2})}{(AB+C^{2})^{2}+C^{2}(A-B)^{2}}\right]
\label{eq:cross-section-nunu-phiphi}
\end{multline}
where 
\begin{align}
A & =s-M_{R}^{2}+\Gamma_{R}^{2}\slash4,\nonumber \\
B & =t-M_{R}^{2}+\Gamma_{R}^{2}\slash4,\nonumber \\
C & =\Gamma_{R}M_{R},
\end{align}
and the decay rate of the RH neutrino is approximated by
$\Gamma_{R}\approx\left|Y^{\nu}\right|^{2}M_{R}/16\pi.$ The thermally
averaged cross section for massless particles to the CM cross section
can be obtained as~\cite{Cannoni:2013bza} 
\begin{equation}
\left<\sigma v\right>_{0}=\dfrac{1}{32T^{5}}\int_{0}^{\infty}ds\, s^{3/2}K_{1}(\sqrt{s}\slash T)\sigma_{CM}(s).
\end{equation}

For the $\nu_{L}\nu_{L}\leftrightarrow\phi\phi$ process, the thermally
averaged cross section can be approximated as 
\begin{equation}
\left\langle \sigma\left(\nu_{L}\nu_{L}\leftrightarrow\phi\phi\right)v\right\rangle _{0}\approx\frac{\left|Y^{\nu}\right|^{4}}{16\pi M_{R}^{2}},
\end{equation}
in the limit  $T\ll M_{R}$. 
To account for the $\sim 0.1 \,\text{eV}$ left-handed neutrino mass, the sum from two channels gives about $\left\langle \sigma v\right\rangle _{0}\sim 5 \times 10^{-30}\,\text{GeV}^{-2}$
for the EGH scenario.

\subsection{Boltzmann Transport Equation}

The lepton number violating processes described above are usually not
fast enough to reach chemical equilibrium, due to the suppression
from the large Majorana mass. Nevertheless, the relaxation of the
system toward its equilibrium can be described by the Boltzmann transport
equation. To the first order in $\mu_{0}/T$, we have 
\begin{align}
\dot{n}_{L}+3Hn_{L} & =-2n_{0}^{eq}\left\langle \sigma v\right\rangle _{0}\left(n_{L}-\dfrac{2\mu_{0}}{T}n_{0}^{eq}\right)\label{eq:nL}
\end{align}
where $n_{L}=n_{\nu}-n_{\overline{\nu}}$ is the total asymmetry of
neutrinos, and $n_{0}^{eq}=T^{3}\slash\pi^{2}$. If the interaction
were fast enough, the system would yeld the lepton asymmetry $n_{L,\, eq}=\frac{2}{\pi^{2}}\mu_{0}T^{2}$
in equilibrium.  The derivation of this equation is discussed in Ref.~\cite{Yang:2015ida}.

\subsection{Resulting Lepton Asymmetry}

As in \cite{Kusenko:2014lra}, the evolution of the lepton asymmetry can be analyzed in two regimes: during the relaxation of the $\sigma$ field ($\mu_{0}\neq0$) and during the subsequent cooling of the Universe ($\mu_{0}\approx0$). During the relaxation of the $\sigma$ field, when $\mu_{0}\propto\partial_{0}\sigma^{2}\neq0$, the Universe produces most of the lepton asymmetry; as mentioned this time is generally insufficient for the system to reach equilibrium, and so the asymmetry produced is of the order $n_{L,\mathrm{eq}} \times \sigma_R T_\mathrm{rlx}^3 t_{\mathrm{rlx}}$, where $t_{\mathrm{rlx}}$ is the time period during which the Higgs field relaxes, and $T_\mathrm{rlx} = T(t_\mathrm{rlx})$. The equilibrium lepton asymmetry can be approximated by its value at $t_{\text{rlx}}$ as 
\begin{equation}
n_{L,\mathrm{eq}}\sim\frac{2}{\pi^{2}}\mu_{0}T_{\text{rlx}}^{2}=\frac{2}{\pi^{2}}\frac{\partial_{0}\sigma^{2}}{M_{n}^{2}}T_{\text{rlx}}^{2}\sim\frac{2}{\pi^{2}}\frac{\sigma_{0}^{2}}{M_{n}^{2}t_{\text{rlx}}}T_{\text{rlx}}^{2},
\end{equation}
where $\sigma_{0}$ is the initial VEV of $\sigma$ given by Eq.~\eqref{eq:initial_VEV}, and $M_{n}$ is the new scale. From this, we find the approximate number density
\begin{equation}
n_{L,\,\text{rlx}}\sim\frac{2\sigma_{0}^{2}T_{\text{rlx}}^{2}}{\pi^{2}M_{n}^{2}t_{\text{rlx}}}\text{min}\left\{ 1,\,\frac{2}{\pi^{2}}\sigma_{R}T_{\text{rlx}}^{3}t_{\text{rlx}}\right\} ,
\end{equation}
where $\sigma_{R}$ is the thermally averaged cross section given by Eq.~\eqref{eq:sigma_R}. 
%Note for the case that the relaxation time is less than the reheating time scale, $t_{\text{rlx}} < t_{R} \equiv 1/\Gamma_{I}$, the temperature at $t_{\text{rlx}}$ can be computed using $T\left(t\right) \approx T_{R}\left(t_{R}/t\right)^{1/4}.$ 

Note the temperature before and after the reheating completes ($t_R\equiv 1/\Gamma_{I}$) is given by
\begin{equation}
T\left(t\right) \approx 
\begin{cases}
T_{R}\left(t_R / t \right)^{1/4} & t < t_{R}\\
\left(\frac{45}{16\pi^{3}g_{*}}\right)^{1/4}\sqrt{M_{pl} / t} & t > t_{R}.
\end{cases}
\end{equation}
We note that because we consider a pure $\phi^4$ potential, the relaxation time is $t_{\text{rlx}}\approx7/\sqrt{\lambda_{\text{eff}}}\sigma_{0}$.  (This holds within a factor of two even for quartic couplings of the order $ \lambda \sim 10^{-20}$.)

After the initial relaxation, the produced asymmetry can be partially washed out.  This is due either to subsequent oscillations, during which the sign of the chemical potential oscillates, or due to ongoing lepton-number-violating interactions in the plasma after the chemical reaction has become small. As computed in \cite{Kusenko:2014lra, Kusenko:2016vcq}, the effect of washout can be approximated by

\begin{align}
N_{L,\, f} & = N_{L,\,\mathrm{rlx}}\exp\left\{ -\frac{\sigma_{R}T_{R}^{3}}{\pi^{2}\Gamma_{I}}\left[8\left(1-\frac{T_{R}}{T_{\mathrm{rlx}}}\right)+\sqrt{15}\right]\right\} \\
 & \approx N_{L,\,\mathrm{rlx}}\exp\left(-\frac{8+\sqrt{15}}{\pi^{2}}\frac{\sigma_{R}T_{R}^{3}}{\Gamma_{I}}\right)
\end{align}
for $t_{\mathrm{rlx}} < t_{R}$, and 
\begin{equation}
N_{L,\, f} = N_{L,\,\mathrm{rlx}}\exp\left(-\frac{\sqrt{15}}{\pi^{2}}\frac{\sigma_{R}T_{R}^{2}T_{\mathrm{rlx}}}{\Gamma_{I}}\right)
\end{equation}
for $t_{\mathrm{rlx}} > t_{R}$.
We note that depending on the strength of the lepton-number-violating potential, it may be advantageous to arrange for a comparatively rapid decay of the $\sigma$ condensate, so that the oscillation amplitude of the scalar VEV is significantly damped.

The asymmetry is also further diluted by $\left(a_{\text{rlx}}/a_{R}\right)^{3}\approx t_{\text{rlx}}^{2}\Gamma_{I}^{2}$ for the case that $t_\mathrm{rlx} < t_R$. Thus at the end of reheating (assuming the oscillation of the scalar field has ended), the Universe obtains the the ratio of the lepton asymmetry to entropy 
\begin{align}
Y & \equiv \frac{n_L}{s} = \frac{45}{2\pi^{2}g_{*S}}\frac{n_{L}}{T^{3}}\nonumber \\
 & \sim \frac{45}{2\pi^{2}g_{*S}} \frac{2\sigma_{0}^{2}}{\pi^{2}M_{n}^{2}} \frac{T_{\text{rlx}}^{2}t_{\text{rlx}}\Gamma_{I}^{2}}{T_{R}^{3}}
 \text{min}\left\{ 1,\:\frac{2}{\pi^{2}}\sigma_{R}T_{\text{rlx}}^{3}t_{\text{rlx}}\right\}
 \exp\left(-\frac{8+\sqrt{15}}{\pi^{2}}\frac{\sigma_{R}T_{R}^{3}}{\Gamma_{I}}\right)
 \label{eq:estimation_formula}
\end{align}
for $t_{\mathrm{rlx}} < t_R$, and 
\begin{equation}
Y \sim\frac{45}{2\pi^{2}g_{*S}}\frac{2\sigma_{0}^{2}}{\pi^{2}M_{n}^{2}}\frac{1}{T_{\mathrm{rlx}}t_{\mathrm{rlx}}}\mathrm{min}\left\{ 1,\frac{2}{\pi^{2}}\sigma_{R}T_{\mathrm{rlx}}^{3}t_{\mathrm{rlx}}\right\} \exp\left(-\frac{\sqrt{15}}{\pi^{2}}\frac{\sigma_{R}T_{R}^{2}T_{\mathrm{rlx}}}{\Gamma_{I}}\right)
\label{eq:estimation_formula-late}
\end{equation}
for $t_{\mathrm{rlx}} > t_R$.
This estimation formula agrees within one order of magnitude with
the numerical result.

%\begin{figure}
%\begin{centering}
%\includegraphics[width=0.6\columnwidth]{Single_lambda_v2}
%\par\end{centering}

%\protect\caption{Evolution of the lepton asymmetry $Y = n_{L}/\left(2\pi^{2}g_{*S}T^{3}/45\right)$ for $\lambda_{\text{eff}} = 10^{-10}$, with the parameters $\Lambda_{I} = 8 \times 10^{15}\:\text{GeV}$, $\Gamma_{I} = 3 \times 10^{6}\:\text{GeV}$, $M_{n} = 4.2 \times 10^{12}\:\text{GeV}$, and $M_{R} = 4.2 \times 10^{14}\:\text{GeV}$. The initial VEV of the $\sigma$ field is $\sigma_{0} = 6.8 \times 10^{12}\:\text{GeV}.$ The maximum temperature during reheating is $T_{\text{max}} = 4.2 \times 10^{13}\:\text{GeV}$. The first vertical dashed lines denote the time of maximum reheating, the first time the $\sigma$ VEV crosses zero, and the beginning of the radiation-dominated era, form left to right.
%\label{fig:asymmetry_history}}
%\end{figure}

\begin{figure}
\begin{centering}
\includegraphics[width=0.6\columnwidth]{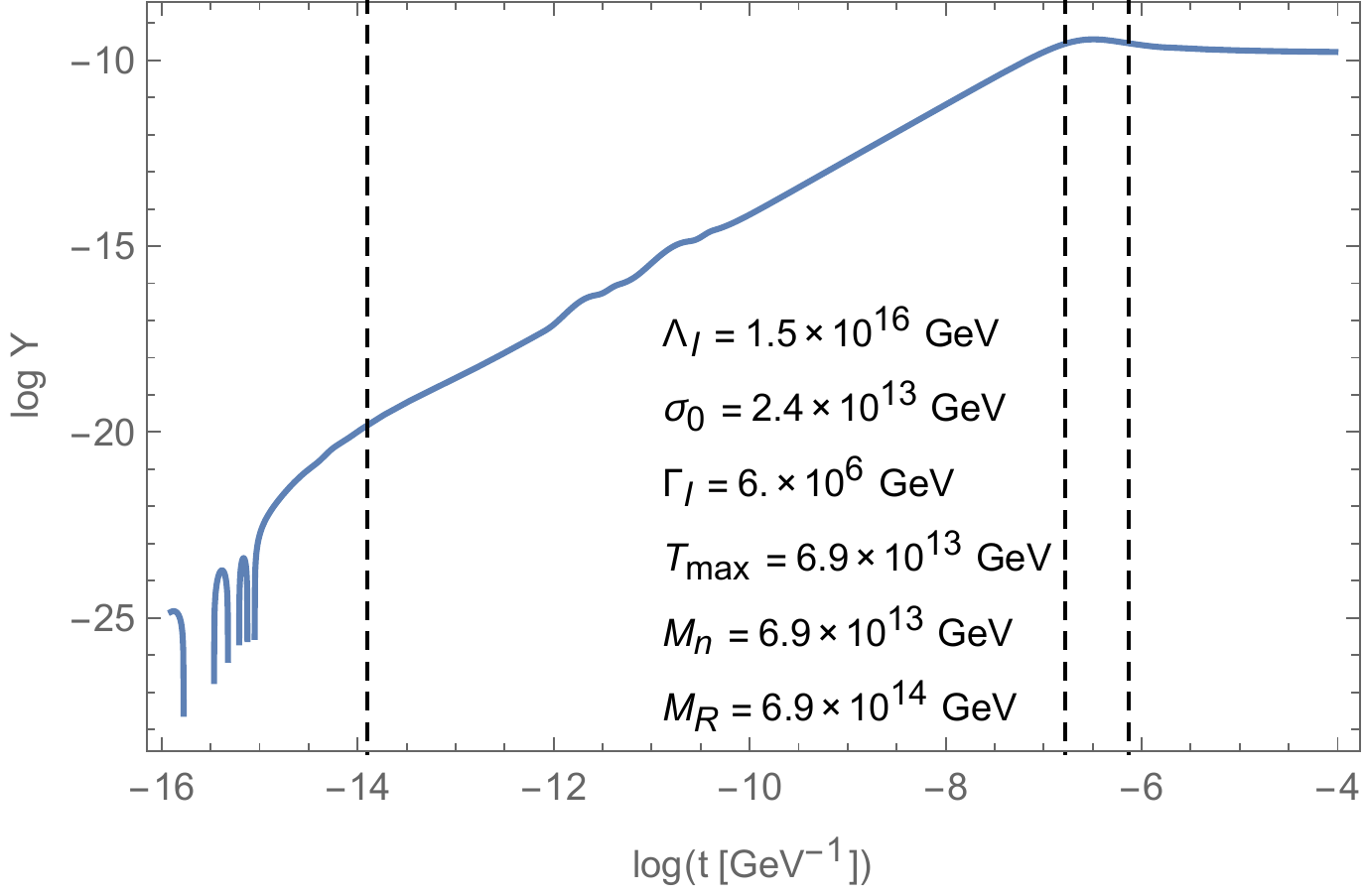}
\par\end{centering}

\protect\caption{Evolution of the lepton asymmetry $Y = n_{L}/\left(2\pi^{2}g_{*S}T^{3}/45\right)$
for $\lambda_{\text{eff}} = 10^{-13}$, with the parameters $\Lambda_{I} = 1.5 \times 10^{16}\:\text{GeV}$,
$\Gamma_{I} = 6 \times 10^{6}\:\text{GeV}$, $M_{n} = 6.9 \times 10^{13}\:\text{GeV}$,
and $M_{R} = 6.9 \times 10^{14}\:\text{GeV}$.
The initial VEV of the $\sigma$ field is $\sigma_{0} = 2.4 \times 10^{13}\:\text{GeV}.$
The maximum temperature during reheating is $T_{\mathrm{max}} = 6.9 \times 10^{13}\:\text{GeV}$.
The vertical dashed lines denote the time of maximum reheating, the beginning of
the radiation-dominated era, and the first time the $\sigma$ VEV crosses zero, form left to right.
\label{fig:asymmetry_history}}
\end{figure}

One can obtain the evolution of the lepton asymmetry more precisely
by solving Eq.~\eqref{eq:nL} numerically. In Fig.~\ref{fig:asymmetry_history},
we present a numerical example for $\lambda_{\text{eff}} = 10^{-13}$
with the inflationary parameters $\Lambda_{I} = 1.5 \times 10^{16}\:\text{GeV}$,
and $\Gamma_{I} = 6 \times 10^{6}\:\text{GeV}$. The RH neutrino
mass scale is set at $M_{R} = 10T_{\text{max}}$ to suppress the thermal
production of RH neutrinos during reheating, and we verify that the resulting
neutrino Yukawa coupling is within the perturbative regime ($\left|Y^{\nu}\right|^{2}/4\pi = 0.18 < 1$). This example gives a late time asymptotic asymmetry about $Y \sim 10^{-10}$.

\begin{figure}
\begin{centering}
\includegraphics[width=0.6\columnwidth]{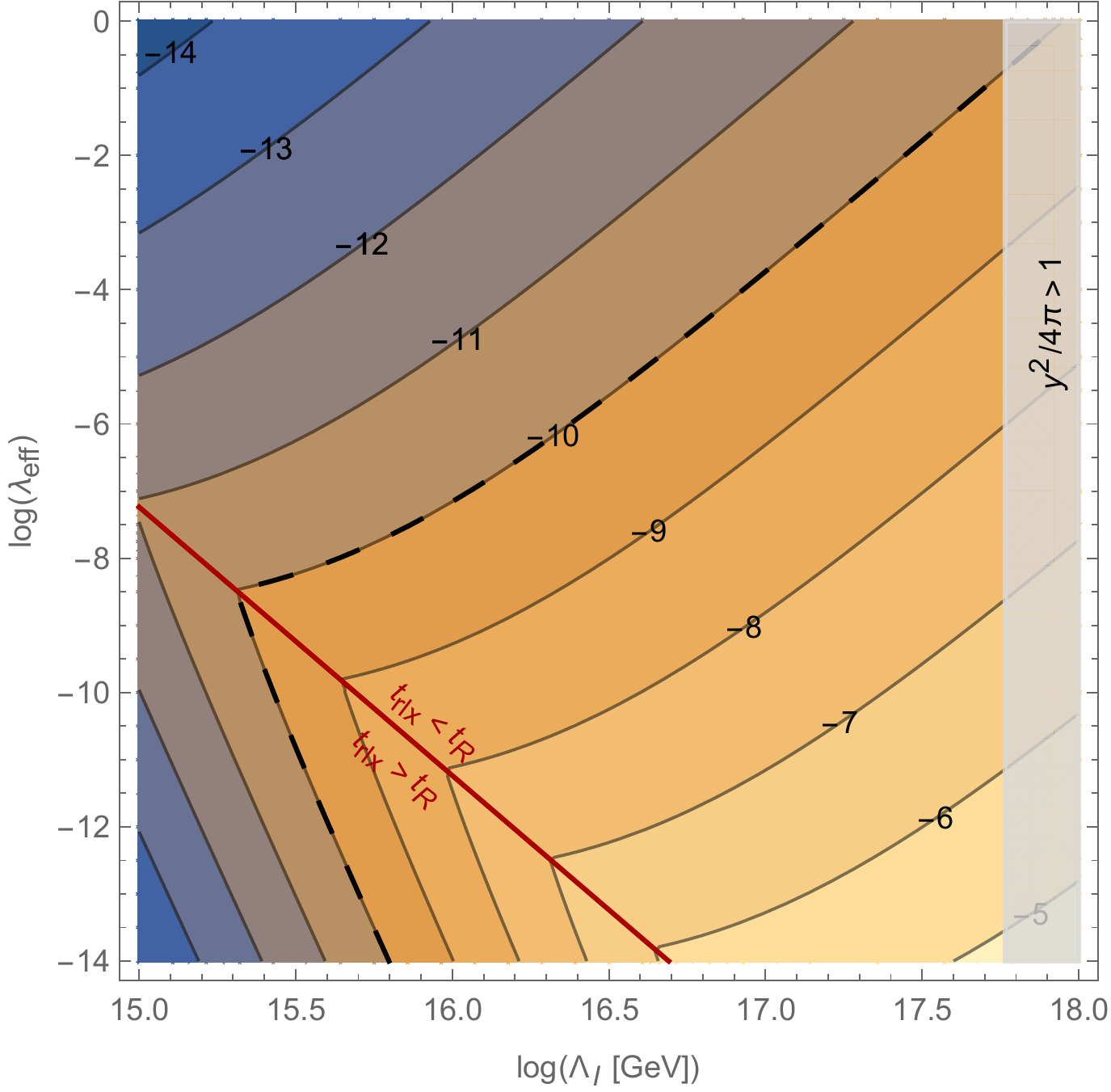}
\par\end{centering}

\caption{The approximate late time asymptotic asymmetry ($\log\left|Y\right|$) using Eqs.~\eqref{eq:estimation_formula} and \eqref{eq:estimation_formula-late} with $\Gamma_{I}=3 \times 10^{6}\,\text{GeV}$, $M_{R}=10 T_{\text{max}}$, and $M_{n}=0.1T_{\text{max}}$. The red line indicated where $t_\mathrm{rlx} = t_R$.}
\label{fig:parameter_space}
\end{figure}

We explored the parameter space of the model using the approximate
formula (Eqs.~\eqref{eq:estimation_formula} and \eqref{eq:estimation_formula-late}). In Fig.~\ref{fig:parameter_space}, we show the approximate late time asymmetry ($\log\left|Y\right|$) as a function of $\lambda_{\text{eff}}$ and $\Lambda_{I}$ with mass scales $M_{R}=10T_{\text{max}}$ and $M_{n}=0.1 T_{\text{max}}$. We have checked that this value of $M_R$ is sufficient to ensure that standard leptogenesis does not produce a sufficiently large asymmetry. The decay rate of the inflaton is set to $\Gamma_{I} = 3.7 \times 10^{6}\:\text{GeV}$ so that it gives the maximum asymmetry for each $\lambda_{\text{eff}}$ and $\Lambda_{I}$. The requirement of the neutrino coupling being perturbative ($\left|Y^{\nu}\right|^{2}/4\pi < 1$) imposes an upper bound on the inflationary energy scale $\Lambda_{I}\lesssim 7 \times 10^{17}\,\text{GeV}$, which is weaker than the bound $\Lambda_I \lesssim 1-2 \times 10^{16} \, \mathrm{GeV}$ from CMB observations \cite{Ade:2013uln,*Ade:2015lrj}.

In Fig.~\ref{fig:parameter_space} we see that the generated asymmetry increases as $\lambda_{\text{eff}}$ decreases in the region where $t_\mathrm{rlx} < t_R$.  This can be understood as follows: the initial VEV is set by inflationary parameters, since by construction, the VEV is permitted only to grow during the last $N \approx 8$ e-folds of inflation.  The relaxation time though scales as $1 \slash \sqrt{ \lambda_{\text{eff}}}$; therefore, as $\lambda_{\text{eff}}$ decreases the system has more time to approach the equilibrium asymmetry.  

The dashed line in the figure indicates $Y = 10^{-10}$, which matches the observed value.  The region of most interest is the lower left side, in which a sufficiently large asymmetry is generated with $\Lambda_I \lesssim 10^{16} \, \mathrm{GeV}$, which is required by observations of the tensor-to-scalar ratio in the CMB \cite{Ade:2013uln,*Ade:2015lrj}.

From this figure, we see the effective quartic coupling is restricted to $\lambda_{\text{eff}}\lesssim 10^{-8}$, which corresponds to a relatively flat effective potential for the $\sigma$ field.  Such flat directions can develop in models with multi-field scalar sectors, such as the EGH model considered here. 

We note that in all of this parameter space, $M_n < T_\mathrm{max}$, and in much of it, $M_n < \sigma_0$.  Because $M_n$ is not the largest scale in the analysis, the use of effective field theory to derive Eq.~\eqref{eq:operator} is questionable.  Consequently, the exact asymmetry would depend on the details of the UV-complete theory considered; however, Eq.~\eqref{eq:operator} is likely to give a reasonable approximation.

\begin{figure}
\includegraphics[scale=.60]{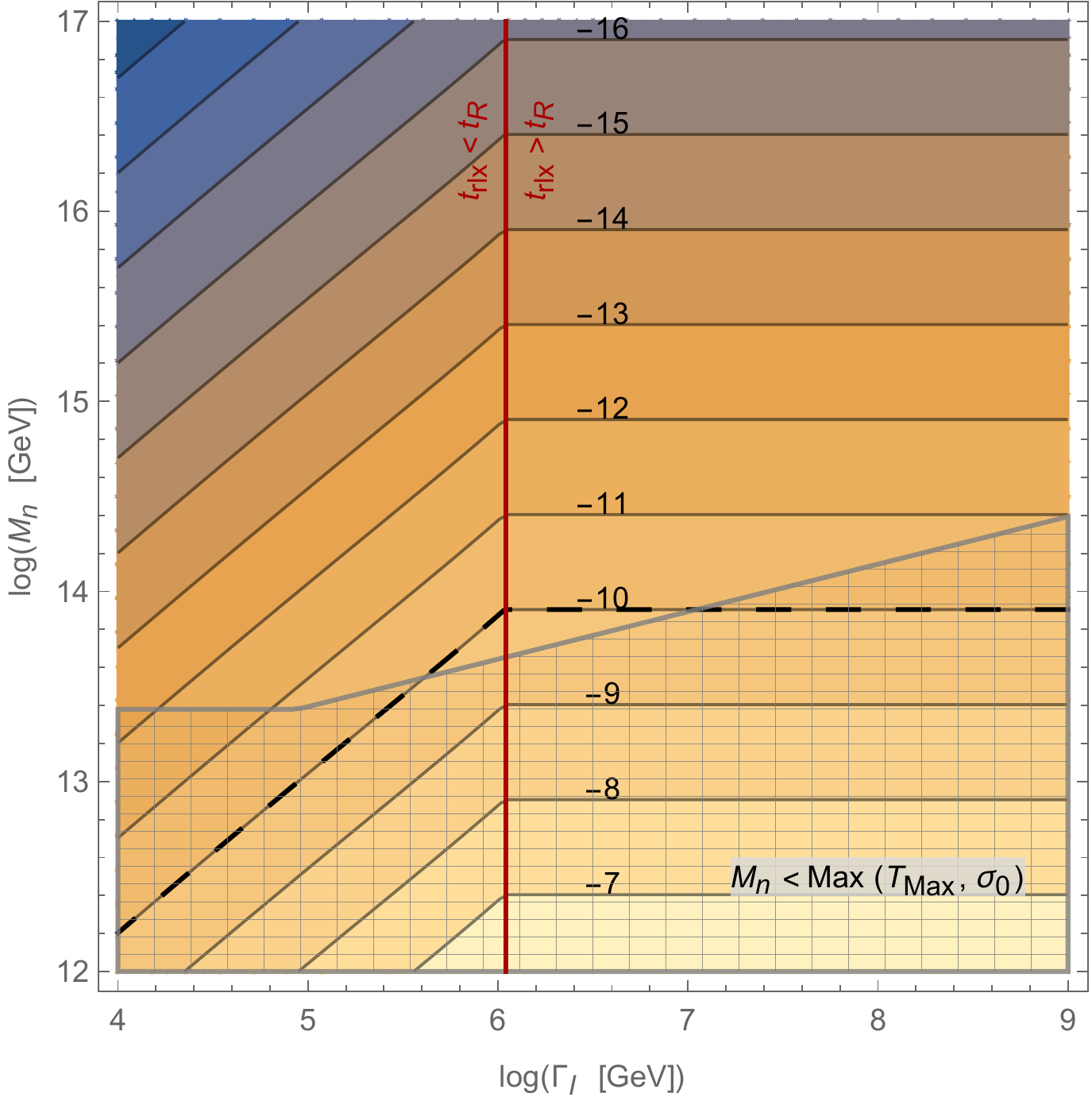}
\includegraphics[scale=.60]{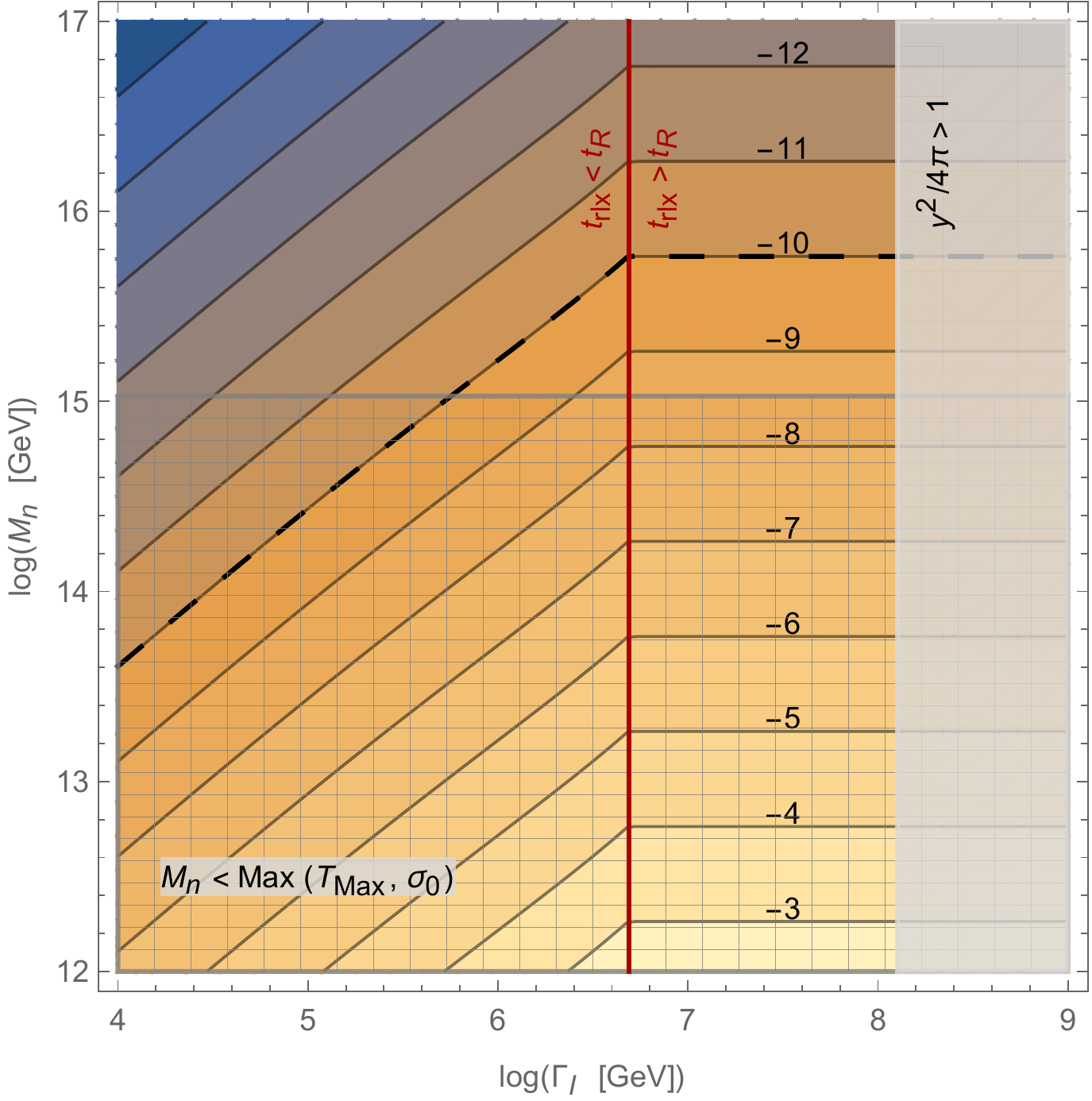}
\caption{An exploration of parameter space as a function of $M_n$ and $\Gamma_I$, for $\Lambda_I = 1.5 \times 10^{16} \; \mathrm{GeV}$ and $\lambda_\mathrm{eff} = 10^{-13}$ (left) and $\Lambda_I = 10^{17} \; \mathrm{GeV}$ and $\lambda_\mathrm{eff} = 10^{-15}$ (right).  The grey area on the right indicates where the theory is not under perturbative control, and the cross hatched region denotes the regime in which the effective field theory used to derive Eq.~\eqref{eq:operator} is unreliable. The red lines indicate where $t_\mathrm{rlx} = t_R$. }
\label{fig:MnVSGammaI}
\end{figure}

In order to compare with our earlier work using the SM Higgs boson, it is convenient to consider the asymmetry as a function of the new scale $M_n$ and the decay rate of the inflaton, $\Gamma_I$.  This is shown in Fig.~\ref{fig:MnVSGammaI}, where we have fixed the inflationary scale $\Lambda_I$ and coupling $\lambda_\mathrm{eff}$.  As above, we also take $M_R = 10 T_\mathrm{max}$.  Therefore, the inflaton decay rate $\Gamma_I$, which fixes $T_\mathrm{max}$, also effectively fixes $M_R$ and therefore, when combined with observational limits on the left-handed neutrino masses, the neutrino coupling $Y^{\nu}$.  Demanding that the coupling $Y^{\nu}$ be in the perturbative regime eliminates the parameter space shown in grey on the right-hand side of the plots.

In these plots, we additionally illustrate the regime in which $\mathrm{Max}(\sigma_0, T_\mathrm{max}) > M_n$ using cross-hatching.  As noted above, in this regime the exact asymmetry would depend on the details of the UV-complete theory considered; however, Eq.~\eqref{eq:operator} is a reasonable estimate.  We see that if one fine-tunes $\lambda_\mathrm{eff}$ to extremely small values ($\mathcal{O}(10^{-13})$ or smaller), an asymmetry $Y \sim 10^{-10}$ can be generated in the regime in which effective field theory is reliable.  For $\Lambda_I = 1.5 \times 10^{16} \, \mathrm{GeV}$, near the upper limit allowed by CMB observations, this region is near $M_n \sim 10^{14} \, \mathrm{GeV}$ and $\Gamma_I \sim 10^6 \, \mathrm{GeV}$.  If $\Lambda_I$ is increased to $10^{17} \, \mathrm{GeV}$ (and the quartic coupling decreased), then the parameter space is significantly larger, as illustrated by the plot on the right.
 For so small values of $\lambda_\mathrm{eff}$ the value of $\theta$ is also small and hence the value of the vev $f$ is large. In fact for $\lambda_\mathrm{eff}=10^{-13}$ the mode of $\theta$ is $\bar \theta = 2.216_{-0.200}^{+0.704}\cdot 10^{-13}$ (the plot on the left) and for $\lambda_\mathrm{eff}=10^{-15}$ (the plot on the right) the mode of $\theta$ is $\bar \theta = 8.359_{-0.487}^{+2.621}\cdot 10^{-8}$.

The plots in Fig.~\ref{fig:MnVSGammaI} illustrate the freedom available in this model.  While the quartic coupling of the SM has a minimum value set by the recently observed Higgs boson, no such constraint restricts the effective EGH self-coupling $\lambda_\mathrm{eff}$ yet.  By decreasing $\lambda_\mathrm{eff}$ one can enhance the asymmetry; alternatively, one may say that lower inflationary scales $\Lambda_I$ and $\Gamma_I$ are permitted in the EGH model.  Furthermore, such small values of $\lambda_\mathrm{eff}$ allow us to generate a sufficiently large asymmetry in the regime in which effective field theory is reliable; in order to accomplish this withing the SM Higgs boson, significant modifications to the Higgs potential (using non-renormalizable operators) at large scales were necessary~\cite{Yang:2015ida}.

As noted above, at small couplings we must be concerned about corrections to the potential.  We note that using the full running coupling \eqref{eq:lambda_eff}
alters the final asymmetry at the level of about 1\%.  One may also be concerned that finite temperature corrections could significantly affect the potential.  However, the finite temperature corrections to the potential scale as $\lambda_\mathrm{eff} T^2 \sigma^2$, and so are also suppressed by the small quartic coupling. For the parameters in Fig.~\ref{fig:asymmetry_history}, this correction only affects the asymmetry by 0.1\%. Couplings to the Standard Model fermions are suppressed by the $\sin(\theta)$ factor, which as indicated by equation \eqref{eq:lambda_eff}, is necessarily small when the quartic coupling is small.

As mentioned above, we can only tune the quartic coupling $\lambda_\mathrm{eff}$ to be this small in models with an extended Higgs sector, as the observed Higgs boson mass suggests a quartic coupling $\mathcal{O}(10^{-2})$ at high scales.  Therefore, we have shown that in an extension of the SM with an extended Higgs sector (included but not limited to the EGH paradigm) it is possible for Higgs-relaxation leptogenesis to generate the observed baryonic asymmetry, although it requires a small self-coupling if the potential in the direction of the VEV has a $\phi^4$ form (and one wishes to remain in the regime in which effective field theory is reliable).  Furthermore, although the parameter space is not large, this can be accomplished for inflationary scales consistent with CMB measurements.

\section{Conclusions and Outlook}

In this work, we have successfully extended the baryogenesis scenario using the relaxation mechanism  to the  EGH framework. Our results show that if the electroweak scale is not fundamental but radiatively generated and consequently the Higgs particle a quasi elementary Goldstone Boson it is possible to generate the baryon asymmetry by marrying the EGH framework to the relaxation mechanism.   In particular, we showed that in order to accommodate baryogenesis the only necessary extensions to the model is to include RH neutrinos and furthermore the operator in \eqref{eq:operator} is generated.   
 
 Because of the nature of the EGH new Higgs sector one can consider very flat scalar potential directions along which the relaxation mechanism can be implemented. This  further translates into achieving a wider region of applicability of the approximations, particularly regarding the regime in which the use effective theory used to derive the \eqref{eq:operator}, as compared to the SM Higgs case.  Specifically baryogenesis can be achieved even from an unmodified $\phi^4$ potential and within a regime in which the effective field theory interpretation of \eqref{eq:operator} is justified, unlike in the SM case.  Observed limits on the inflationary scale $\Lambda_I$ restrict, but do not eliminate, this parameter space.

\section*{Acknowledgements}

The authors wish to thank A.\ Kusenko for helpful discussions. L.Y.\ was supported by the U.S.\ Department of Energy Grant DE-SC0009937.  The work of H.G.\ and F.S.\ is partially supported by the Danish National Research Foundation grant DNRF:90.

\appendix

%Edited by Louis on 2016-03-20
\section{Lepton-Number-Violating Cross Section in the EGH Model\label{app:16}}

In section \ref{sub:Lepton-Number-Violating}, we remarked that the
thermally averaged cross section of the lepton-number-violating processes
in the EGH model is enhanced by a factor of 16 in comparison to the
SM Higgs case.  In this appendix, we discuss the calculation of the cross section, including this prefactor.  The lepton-number-violating
processes that we consider are (i) $\nu_{L}\phi_{i}\leftrightarrow\bar{\nu}_{L}\phi_{j}$
and (ii) $\nu_{L}\nu_{L}\leftrightarrow\phi_{i}\phi_{j}$ as shown
in Fig.~\ref{fig:lepton_violation}. The scalar fields that are involved 
in these processes are $\phi_{i} = \left\{ \sigma,\,\Pi_{4},\,\Theta,\,\tilde{\Pi}_{4},\,\Pi_{3},\,\tilde{\Pi}_{3}\right\} $.
Their cross sections depend on the Yukawa couplings to LH and RH neutrinos,
which can be read from Eq.~\eqref{eq:neutrino_Y} as
\begin{equation}
Y_{i}^{\nu} = \left\{ Y^{\nu}\sin\theta,\, Y^{\nu}\cos\theta,\, iY^{\nu}\sin\theta,\, iY^{\nu}\cos\theta,\, iY^{\nu},\, Y^{\nu}\right\} ,
\label{eq:couplings}
\end{equation}
respectively.

For the (i) $\nu_{L}\phi_{i}\leftrightarrow\bar{\nu}_{L}\phi_{j}$
channel, the cross section depends on $Y_{i}^{\nu}$ as $\sigma\left(\nu_{L}\phi_{i}\rightarrow\bar{\nu}_{L}\phi_{j}\right)\propto\left|Y_{i}^{\nu}\right|^{2}\left|Y_{j}^{\nu}\right|^{2}.$
Summing the contribution from each $i$ and $j$, the thermally averaged
cross section from the channel (i) is 
\begin{equation}
\left\langle \sigma v\right\rangle _{\left(\mathrm{i}\right)}
= \sum_{i,j}\left\langle \sigma\left(\nu_{L}\phi_{i}\leftrightarrow\bar{\nu}_{L}\phi_{j}\right)v\right\rangle 
= \frac{\left\langle \sigma\left(\nu_{L}\phi\leftrightarrow\bar{\nu}_{L}\phi\right)v\right\rangle }{\left|Y^{\nu}\right|^{4}}\sum_{i,j}\left|Y_{i}^{\nu}\right|^{2}\left|Y_{j}^{\nu}\right|^{2}=16\left\langle \sigma\left(\nu_{L}\phi\leftrightarrow\bar{\nu}_{L}\phi\right)v\right\rangle ,
\end{equation}
where $\phi$ is a Standard Model-like scalar with the Yukawa coupling $\left|Y^{\nu}\right|$
to LH and RH neutrinos, and $\left\langle \sigma\left(\nu_{L}\phi\leftrightarrow\bar{\nu}_{L}\phi\right)v\right\rangle $
can be obtained from Eq.~\eqref{eq:cross-section-nunu-phiphi}.

For the (ii) $\nu_{L}\nu_{L}\leftrightarrow\phi_{i}\phi_{j}$ channel,
one has to consider the symmetry factor due to the identical outgoing
particles. The CM cross section in the limit $T\ll M_{R}$ is approximated
as 
\begin{equation}
\sigma\left(\nu_{L}\nu_{L}\rightarrow\phi_{i}\phi_{j}\right) \approx \frac{1}{4\pi S} \frac{\left|Y_{i}^{\nu}\right|^{2}\left|Y_{j}^{\nu}\right|^{2}}{M_{R}^{2}},
\end{equation}
where the symmetry factor $S=2$ if $i=j$, and $S=1$ if $i\neq j$.
Summing different outgoing particles, the net CM cross section is
\begin{align}
\sum_{i}\sum_{j\geq i} 
 \sigma\left(\nu_{L}\nu_{L}\rightarrow\phi_{i}\phi_{j}\right)
 & = \sum_{i}
 \left[
 \sigma\left(\nu_{L}\nu_{L}\rightarrow\phi_{i}\phi_{i}\right)
 + \sum_{j>i}\sigma\left(\nu_{L}\nu_{L}\rightarrow\phi_{i}\phi_{j}\right)
 \right]\\
 & = \frac{\sigma\left(\nu_{L}\nu_{L}\rightarrow\phi\phi\right)}{\left|Y^{\nu}\right|^{4}} 
 \sum_{i}
  \left(
  \left|Y_{i}^{\nu}\right|^{4}
  + 2 \sum_{j>i} \left|Y_{i}^{\nu}\right|^{2}
    \left|Y_{j}^{\nu}\right|^{2}\right)\\
 & = \frac{\sigma\left(\nu_{L}\nu_{L}\rightarrow\phi\phi\right)}{\left|Y^{\nu}\right|^{4}} 
 \sum_{i,j} \left|Y_{i}^{\nu}\right|^{2} \left|Y_{j}^{\nu}\right|^{2}\\
 & = 16 \, \sigma\left(\nu_{L}\nu_{L}\rightarrow\phi\phi\right),
\end{align}
where again $\phi$ denotes a Standard-Model like scalar which has a coupling constant $Y^\nu$ (which is related to the couplings of the EGH scalars via equation \eqref{eq:couplings}).  Again, we see the enhancement by the factor of 16,
\begin{equation}
\left\langle \sigma v\right\rangle _{\mathrm{\left(ii\right)}}
= \sum_{i} \sum_{j\geq i} \left\langle
 \sigma\left(\nu_{L}\nu_{L}\leftrightarrow\phi_{i}\phi_{j}\right)v
 \right\rangle 
= 16 \left\langle \sigma\left(\nu_{L}\nu_{L}\leftrightarrow\phi\phi\right)v \right\rangle .
\end{equation}

\bibliographystyle{apsrev4-1}
\bibliography{Note_Ref}

%merlin.mbs apsrev4-1.bst 2010-07-25 4.21a (PWD, AO, DPC) hacked
%Control: key (0)
%Control: author (72) initials jnrlst
%Control: editor formatted (1) identically to author
%Control: production of article title (-1) disabled
%Control: page (0) single
%Control: year (1) truncated
%Control: production of eprint (0) enabled
\begin{thebibliography}{50}%
\makeatletter
\providecommand \@ifxundefined [1]{%
 \@ifx{#1\undefined}
}%
\providecommand \@ifnum [1]{%
 \ifnum #1\expandafter \@firstoftwo
 \else \expandafter \@secondoftwo
 \fi
}%
\providecommand \@ifx [1]{%
 \ifx #1\expandafter \@firstoftwo
 \else \expandafter \@secondoftwo
 \fi
}%
\providecommand \natexlab [1]{#1}%
\providecommand \enquote  [1]{``#1''}%
\providecommand \bibnamefont  [1]{#1}%
\providecommand \bibfnamefont [1]{#1}%
\providecommand \citenamefont [1]{#1}%
\providecommand \href@noop [0]{\@secondoftwo}%
\providecommand \href [0]{\begingroup \@sanitize@url \@href}%
\providecommand \@href[1]{\@@startlink{#1}\@@href}%
\providecommand \@@href[1]{\endgroup#1\@@endlink}%
\providecommand \@sanitize@url [0]{\catcode `\\12\catcode `\$12\catcode
  `\&12\catcode `\#12\catcode `\^12\catcode `\_12\catcode `\%12\relax}%
\providecommand \@@startlink[1]{}%
\providecommand \@@endlink[0]{}%
\providecommand \url  [0]{\begingroup\@sanitize@url \@url }%
\providecommand \@url [1]{\endgroup\@href {#1}{\urlprefix }}%
\providecommand \urlprefix  [0]{URL }%
\providecommand \Eprint [0]{\href }%
\providecommand \doibase [0]{http://dx.doi.org/}%
\providecommand \selectlanguage [0]{\@gobble}%
\providecommand \bibinfo  [0]{\@secondoftwo}%
\providecommand \bibfield  [0]{\@secondoftwo}%
\providecommand \translation [1]{[#1]}%
\providecommand \BibitemOpen [0]{}%
\providecommand \bibitemStop [0]{}%
\providecommand \bibitemNoStop [0]{.\EOS\space}%
\providecommand \EOS [0]{\spacefactor3000\relax}%
\providecommand \BibitemShut  [1]{\csname bibitem#1\endcsname}%
\let\auto@bib@innerbib\@empty
%</preamble>
\bibitem [{\citenamefont {Sakharov}(1967)}]{Sakharov:1967dj}%
  \BibitemOpen
  \bibfield  {author} {\bibinfo {author} {\bibfnamefont {A.~D.}\ \bibnamefont
  {Sakharov}},\ }\href {\doibase 10.1070/PU1991v034n05ABEH002497} {\bibfield
  {journal} {\bibinfo  {journal} {Pisma Zh. Eksp. Teor. Fiz.}\ }\textbf
  {\bibinfo {volume} {5}},\ \bibinfo {pages} {32} (\bibinfo {year} {1967})},\
  \bibinfo {note} {[Usp. Fiz. Nauk161,61(1991)]}\BibitemShut {NoStop}%
%%CITATION = ZFPRA,5,32;%%
\bibitem [{\citenamefont {Cline}(2006)}]{Cline:2006ts}%
  \BibitemOpen
  \bibfield  {author} {\bibinfo {author} {\bibfnamefont {J.~M.}\ \bibnamefont
  {Cline}},\ }in\ \href@noop {} {\emph {\bibinfo {booktitle} {{Les Houches
  Summer School - Session 86: Particle Physics and Cosmology: The Fabric of
  Spacetime Les Houches, France, July 31-August 25, 2006}}}}\ (\bibinfo {year}
  {2006})\ \Eprint {http://arxiv.org/abs/hep-ph/0609145} {arXiv:hep-ph/0609145
  [hep-ph]} \BibitemShut {NoStop}%
%%CITATION = HEP-PH/0609145;%%
\bibitem [{\citenamefont {Bunch}\ and\ \citenamefont
  {Davies}(1978)}]{Bunch:1978yq}%
  \BibitemOpen
  \bibfield  {author} {\bibinfo {author} {\bibfnamefont {T.}~\bibnamefont
  {Bunch}}\ and\ \bibinfo {author} {\bibfnamefont {P.}~\bibnamefont {Davies}},\
  }\href {\doibase 10.1098/rspa.1978.0060} {\bibfield  {journal} {\bibinfo
  {journal} {Proc. Roy. Soc. Lond.}\ }\textbf {\bibinfo {volume} {A360}},\
  \bibinfo {pages} {117} (\bibinfo {year} {1978})}\BibitemShut {NoStop}%
%%CITATION = PRSLA,A360,117;%%
\bibitem [{\citenamefont {Linde}(1982)}]{Linde:1982}%
  \BibitemOpen
  \bibfield  {author} {\bibinfo {author} {\bibfnamefont {A.~D.}\ \bibnamefont
  {Linde}},\ }\href {\doibase 10.1016/0370-2693(82)90293-3} {\bibfield
  {journal} {\bibinfo  {journal} {Phys. Lett.}\ }\textbf {\bibinfo {volume}
  {B116}},\ \bibinfo {pages} {335} (\bibinfo {year} {1982})}\BibitemShut
  {NoStop}%
%%CITATION = PHLTA,B116,335;%%
\bibitem [{\citenamefont {Starobinsky}\ and\ \citenamefont
  {Yokoyama}(1994)}]{Starobinsky:1994bd}%
  \BibitemOpen
  \bibfield  {author} {\bibinfo {author} {\bibfnamefont {A.~A.}\ \bibnamefont
  {Starobinsky}}\ and\ \bibinfo {author} {\bibfnamefont {J.}~\bibnamefont
  {Yokoyama}},\ }\href {\doibase 10.1103/PhysRevD.50.6357} {\bibfield
  {journal} {\bibinfo  {journal} {Phys. Rev.}\ }\textbf {\bibinfo {volume}
  {D50}},\ \bibinfo {pages} {6357} (\bibinfo {year} {1994})},\ \Eprint
  {http://arxiv.org/abs/astro-ph/9407016} {arXiv:astro-ph/9407016 [astro-ph]}
  \BibitemShut {NoStop}%
%%CITATION = ASTRO-PH/9407016;%%
\bibitem [{\citenamefont {Cohen}\ and\ \citenamefont
  {Kaplan}(1987)}]{Cohen:1987vi}%
  \BibitemOpen
  \bibfield  {author} {\bibinfo {author} {\bibfnamefont {A.~G.}\ \bibnamefont
  {Cohen}}\ and\ \bibinfo {author} {\bibfnamefont {D.~B.}\ \bibnamefont
  {Kaplan}},\ }\href {\doibase 10.1016/0370-2693(87)91369-4} {\bibfield
  {journal} {\bibinfo  {journal} {Phys. Lett.}\ }\textbf {\bibinfo {volume}
  {B199}},\ \bibinfo {pages} {251} (\bibinfo {year} {1987})}\BibitemShut
  {NoStop}%
%%CITATION = PHLTA,B199,251;%%
\bibitem [{\citenamefont {Dine}\ \emph {et~al.}(1991)\citenamefont {Dine},
  \citenamefont {Huet}, \citenamefont {Singleton},\ and\ \citenamefont
  {Susskind}}]{Dine:1990fj}%
  \BibitemOpen
  \bibfield  {author} {\bibinfo {author} {\bibfnamefont {M.}~\bibnamefont
  {Dine}}, \bibinfo {author} {\bibfnamefont {P.}~\bibnamefont {Huet}}, \bibinfo
  {author} {\bibfnamefont {R.~L.}\ \bibnamefont {Singleton}}, \ and\ \bibinfo
  {author} {\bibfnamefont {L.}~\bibnamefont {Susskind}},\ }\href {\doibase
  10.1016/0370-2693(91)91905-B} {\bibfield  {journal} {\bibinfo  {journal}
  {Phys. Lett.}\ }\textbf {\bibinfo {volume} {B257}},\ \bibinfo {pages} {351}
  (\bibinfo {year} {1991})}\BibitemShut {NoStop}%
%%CITATION = PHLTA,B257,351;%%
\bibitem [{\citenamefont {Kusenko}\ \emph
  {et~al.}(2015{\natexlab{a}})\citenamefont {Kusenko}, \citenamefont {Pearce},\
  and\ \citenamefont {Yang}}]{Kusenko:2014lra}%
  \BibitemOpen
  \bibfield  {author} {\bibinfo {author} {\bibfnamefont {A.}~\bibnamefont
  {Kusenko}}, \bibinfo {author} {\bibfnamefont {L.}~\bibnamefont {Pearce}}, \
  and\ \bibinfo {author} {\bibfnamefont {L.}~\bibnamefont {Yang}},\ }\href
  {\doibase 10.1103/PhysRevLett.114.061302} {\bibfield  {journal} {\bibinfo
  {journal} {Phys. Rev. Lett.}\ }\textbf {\bibinfo {volume} {114}},\ \bibinfo
  {pages} {061302} (\bibinfo {year} {2015}{\natexlab{a}})},\ \Eprint
  {http://arxiv.org/abs/1410.0722} {arXiv:1410.0722 [hep-ph]} \BibitemShut
  {NoStop}%
%%CITATION = ARXIV:1410.0722;%%
\bibitem [{\citenamefont {Pearce}\ \emph {et~al.}(2015)\citenamefont {Pearce},
  \citenamefont {Yang}, \citenamefont {Kusenko},\ and\ \citenamefont
  {Peloso}}]{Pearce:2015nga}%
  \BibitemOpen
  \bibfield  {author} {\bibinfo {author} {\bibfnamefont {L.}~\bibnamefont
  {Pearce}}, \bibinfo {author} {\bibfnamefont {L.}~\bibnamefont {Yang}},
  \bibinfo {author} {\bibfnamefont {A.}~\bibnamefont {Kusenko}}, \ and\
  \bibinfo {author} {\bibfnamefont {M.}~\bibnamefont {Peloso}},\ }\href
  {\doibase 10.1103/PhysRevD.92.023509} {\bibfield  {journal} {\bibinfo
  {journal} {Phys. Rev.}\ }\textbf {\bibinfo {volume} {D92}},\ \bibinfo {pages}
  {023509} (\bibinfo {year} {2015})},\ \Eprint
  {http://arxiv.org/abs/1505.02461} {arXiv:1505.02461 [hep-ph]} \BibitemShut
  {NoStop}%
%%CITATION = ARXIV:1505.02461;%%
\bibitem [{\citenamefont {Yang}\ \emph {et~al.}(2015)\citenamefont {Yang},
  \citenamefont {Pearce},\ and\ \citenamefont {Kusenko}}]{Yang:2015ida}%
  \BibitemOpen
  \bibfield  {author} {\bibinfo {author} {\bibfnamefont {L.}~\bibnamefont
  {Yang}}, \bibinfo {author} {\bibfnamefont {L.}~\bibnamefont {Pearce}}, \ and\
  \bibinfo {author} {\bibfnamefont {A.}~\bibnamefont {Kusenko}},\ }\href
  {\doibase 10.1103/PhysRevD.92.043506} {\bibfield  {journal} {\bibinfo
  {journal} {Phys. Rev.}\ }\textbf {\bibinfo {volume} {D92}},\ \bibinfo {pages}
  {043506} (\bibinfo {year} {2015})},\ \Eprint
  {http://arxiv.org/abs/1505.07912} {arXiv:1505.07912 [hep-ph]} \BibitemShut
  {NoStop}%
%%CITATION = ARXIV:1505.07912;%%
\bibitem [{\citenamefont {Kusenko}\ \emph
  {et~al.}(2015{\natexlab{b}})\citenamefont {Kusenko}, \citenamefont
  {Schmitz},\ and\ \citenamefont {Yanagida}}]{Kusenko:2014uta}%
  \BibitemOpen
  \bibfield  {author} {\bibinfo {author} {\bibfnamefont {A.}~\bibnamefont
  {Kusenko}}, \bibinfo {author} {\bibfnamefont {K.}~\bibnamefont {Schmitz}}, \
  and\ \bibinfo {author} {\bibfnamefont {T.~T.}\ \bibnamefont {Yanagida}},\
  }\href {\doibase 10.1103/PhysRevLett.115.011302} {\bibfield  {journal}
  {\bibinfo  {journal} {Phys. Rev. Lett.}\ }\textbf {\bibinfo {volume} {115}},\
  \bibinfo {pages} {011302} (\bibinfo {year} {2015}{\natexlab{b}})},\ \Eprint
  {http://arxiv.org/abs/1412.2043} {arXiv:1412.2043 [hep-ph]} \BibitemShut
  {NoStop}%
%%CITATION = ARXIV:1412.2043;%%
\bibitem [{\citenamefont {Adshead}\ and\ \citenamefont
  {Sfakianakis}(2015)}]{Adshead:2015jza}%
  \BibitemOpen
  \bibfield  {author} {\bibinfo {author} {\bibfnamefont {P.}~\bibnamefont
  {Adshead}}\ and\ \bibinfo {author} {\bibfnamefont {E.~I.}\ \bibnamefont
  {Sfakianakis}},\ }\href@noop {} {\  (\bibinfo {year} {2015})},\ \Eprint
  {http://arxiv.org/abs/1508.00881} {arXiv:1508.00881 [hep-ph]} \BibitemShut
  {NoStop}%
%%CITATION = ARXIV:1508.00881;%%
\bibitem [{\citenamefont {Li}\ \emph {et~al.}(2002)\citenamefont {Li},
  \citenamefont {Wang}, \citenamefont {Feng},\ and\ \citenamefont
  {Zhang}}]{Li:2001st}%
  \BibitemOpen
  \bibfield  {author} {\bibinfo {author} {\bibfnamefont {M.-z.}\ \bibnamefont
  {Li}}, \bibinfo {author} {\bibfnamefont {X.-l.}\ \bibnamefont {Wang}},
  \bibinfo {author} {\bibfnamefont {B.}~\bibnamefont {Feng}}, \ and\ \bibinfo
  {author} {\bibfnamefont {X.-m.}\ \bibnamefont {Zhang}},\ }\href {\doibase
  10.1103/PhysRevD.65.103511} {\bibfield  {journal} {\bibinfo  {journal} {Phys.
  Rev.}\ }\textbf {\bibinfo {volume} {D65}},\ \bibinfo {pages} {103511}
  (\bibinfo {year} {2002})},\ \Eprint {http://arxiv.org/abs/hep-ph/0112069}
  {arXiv:hep-ph/0112069 [hep-ph]} \BibitemShut {NoStop}%
%%CITATION = HEP-PH/0112069;%%
\bibitem [{\citenamefont {De~Felice}\ \emph {et~al.}(2003)\citenamefont
  {De~Felice}, \citenamefont {Nasri},\ and\ \citenamefont
  {Trodden}}]{DeFelice:2002ir}%
  \BibitemOpen
  \bibfield  {author} {\bibinfo {author} {\bibfnamefont {A.}~\bibnamefont
  {De~Felice}}, \bibinfo {author} {\bibfnamefont {S.}~\bibnamefont {Nasri}}, \
  and\ \bibinfo {author} {\bibfnamefont {M.}~\bibnamefont {Trodden}},\ }\href
  {\doibase 10.1103/PhysRevD.67.043509} {\bibfield  {journal} {\bibinfo
  {journal} {Phys. Rev.}\ }\textbf {\bibinfo {volume} {D67}},\ \bibinfo {pages}
  {043509} (\bibinfo {year} {2003})},\ \Eprint
  {http://arxiv.org/abs/hep-ph/0207211} {arXiv:hep-ph/0207211 [hep-ph]}
  \BibitemShut {NoStop}%
%%CITATION = HEP-PH/0207211;%%
\bibitem [{\citenamefont {Chiba}\ \emph {et~al.}(2004)\citenamefont {Chiba},
  \citenamefont {Takahashi},\ and\ \citenamefont {Yamaguchi}}]{Chiba:2003vp}%
  \BibitemOpen
  \bibfield  {author} {\bibinfo {author} {\bibfnamefont {T.}~\bibnamefont
  {Chiba}}, \bibinfo {author} {\bibfnamefont {F.}~\bibnamefont {Takahashi}}, \
  and\ \bibinfo {author} {\bibfnamefont {M.}~\bibnamefont {Yamaguchi}},\ }\href
  {\doibase 10.1103/PhysRevLett.114.209901, 10.1103/PhysRevLett.92.011301}
  {\bibfield  {journal} {\bibinfo  {journal} {Phys. Rev. Lett.}\ }\textbf
  {\bibinfo {volume} {92}},\ \bibinfo {pages} {011301} (\bibinfo {year}
  {2004})},\ \bibinfo {note} {[Erratum: Phys. Rev.
  Lett.114,no.20,209901(2015)]},\ \Eprint {http://arxiv.org/abs/hep-ph/0304102}
  {arXiv:hep-ph/0304102 [hep-ph]} \BibitemShut {NoStop}%
%%CITATION = HEP-PH/0304102;%%
\bibitem [{\citenamefont {Takahashi}\ and\ \citenamefont
  {Yamaguchi}(2004)}]{Takahashi:2003db}%
  \BibitemOpen
  \bibfield  {author} {\bibinfo {author} {\bibfnamefont {F.}~\bibnamefont
  {Takahashi}}\ and\ \bibinfo {author} {\bibfnamefont {M.}~\bibnamefont
  {Yamaguchi}},\ }\href {\doibase 10.1103/PhysRevD.69.083506} {\bibfield
  {journal} {\bibinfo  {journal} {Phys. Rev.}\ }\textbf {\bibinfo {volume}
  {D69}},\ \bibinfo {pages} {083506} (\bibinfo {year} {2004})},\ \Eprint
  {http://arxiv.org/abs/hep-ph/0308173} {arXiv:hep-ph/0308173 [hep-ph]}
  \BibitemShut {NoStop}%
%%CITATION = HEP-PH/0308173;%%
\bibitem [{\citenamefont {Kamada}\ and\ \citenamefont
  {Yamaguchi}(2012)}]{Kamada:2012ht}%
  \BibitemOpen
  \bibfield  {author} {\bibinfo {author} {\bibfnamefont {K.}~\bibnamefont
  {Kamada}}\ and\ \bibinfo {author} {\bibfnamefont {M.}~\bibnamefont
  {Yamaguchi}},\ }\href {\doibase 10.1103/PhysRevD.85.103530} {\bibfield
  {journal} {\bibinfo  {journal} {Phys. Rev.}\ }\textbf {\bibinfo {volume}
  {D85}},\ \bibinfo {pages} {103530} (\bibinfo {year} {2012})},\ \Eprint
  {http://arxiv.org/abs/1201.2636} {arXiv:1201.2636 [hep-ph]} \BibitemShut
  {NoStop}%
%%CITATION = ARXIV:1201.2636;%%
\bibitem [{\citenamefont {Alanne}\ \emph
  {et~al.}(2015{\natexlab{a}})\citenamefont {Alanne}, \citenamefont {Gertov},
  \citenamefont {Sannino},\ and\ \citenamefont {Tuominen}}]{Alanne:2014kea}%
  \BibitemOpen
  \bibfield  {author} {\bibinfo {author} {\bibfnamefont {T.}~\bibnamefont
  {Alanne}}, \bibinfo {author} {\bibfnamefont {H.}~\bibnamefont {Gertov}},
  \bibinfo {author} {\bibfnamefont {F.}~\bibnamefont {Sannino}}, \ and\
  \bibinfo {author} {\bibfnamefont {K.}~\bibnamefont {Tuominen}},\ }\href
  {\doibase 10.1103/PhysRevD.91.095021} {\bibfield  {journal} {\bibinfo
  {journal} {Phys. Rev.}\ }\textbf {\bibinfo {volume} {D91}},\ \bibinfo {pages}
  {095021} (\bibinfo {year} {2015}{\natexlab{a}})},\ \Eprint
  {http://arxiv.org/abs/1411.6132} {arXiv:1411.6132 [hep-ph]} \BibitemShut
  {NoStop}%
%%CITATION = ARXIV:1411.6132;%%
\bibitem [{\citenamefont {Gertov}\ \emph {et~al.}(2015)\citenamefont {Gertov},
  \citenamefont {Meroni}, \citenamefont {Molinaro},\ and\ \citenamefont
  {Sannino}}]{Gertov:2015xma}%
  \BibitemOpen
  \bibfield  {author} {\bibinfo {author} {\bibfnamefont {H.}~\bibnamefont
  {Gertov}}, \bibinfo {author} {\bibfnamefont {A.}~\bibnamefont {Meroni}},
  \bibinfo {author} {\bibfnamefont {E.}~\bibnamefont {Molinaro}}, \ and\
  \bibinfo {author} {\bibfnamefont {F.}~\bibnamefont {Sannino}},\ }\href@noop
  {} {\  (\bibinfo {year} {2015})},\ \Eprint {http://arxiv.org/abs/1507.06666}
  {arXiv:1507.06666 [hep-ph]} \BibitemShut {NoStop}%
%%CITATION = ARXIV:1507.06666;%%
\bibitem [{\citenamefont {Kaplan}\ and\ \citenamefont
  {Georgi}(1984)}]{Kaplan:1983fs}%
  \BibitemOpen
  \bibfield  {author} {\bibinfo {author} {\bibfnamefont {D.~B.}\ \bibnamefont
  {Kaplan}}\ and\ \bibinfo {author} {\bibfnamefont {H.}~\bibnamefont
  {Georgi}},\ }\href {\doibase 10.1016/0370-2693(84)91177-8} {\bibfield
  {journal} {\bibinfo  {journal} {Phys. Lett.}\ }\textbf {\bibinfo {volume}
  {B136}},\ \bibinfo {pages} {183} (\bibinfo {year} {1984})}\BibitemShut
  {NoStop}%
%%CITATION = PHLTA,B136,183;%%
\bibitem [{\citenamefont {Kaplan}\ \emph {et~al.}(1984)\citenamefont {Kaplan},
  \citenamefont {Georgi},\ and\ \citenamefont {Dimopoulos}}]{Kaplan:1983sm}%
  \BibitemOpen
  \bibfield  {author} {\bibinfo {author} {\bibfnamefont {D.~B.}\ \bibnamefont
  {Kaplan}}, \bibinfo {author} {\bibfnamefont {H.}~\bibnamefont {Georgi}}, \
  and\ \bibinfo {author} {\bibfnamefont {S.}~\bibnamefont {Dimopoulos}},\
  }\href {\doibase 10.1016/0370-2693(84)91178-X} {\bibfield  {journal}
  {\bibinfo  {journal} {Phys. Lett.}\ }\textbf {\bibinfo {volume} {B136}},\
  \bibinfo {pages} {187} (\bibinfo {year} {1984})}\BibitemShut {NoStop}%
%%CITATION = PHLTA,B136,187;%%
\bibitem [{\citenamefont {Cacciapaglia}\ and\ \citenamefont
  {Sannino}(2014)}]{Cacciapaglia:2014uja}%
  \BibitemOpen
  \bibfield  {author} {\bibinfo {author} {\bibfnamefont {G.}~\bibnamefont
  {Cacciapaglia}}\ and\ \bibinfo {author} {\bibfnamefont {F.}~\bibnamefont
  {Sannino}},\ }\href {\doibase 10.1007/JHEP04(2014)111} {\bibfield  {journal}
  {\bibinfo  {journal} {JHEP}\ }\textbf {\bibinfo {volume} {04}},\ \bibinfo
  {pages} {111} (\bibinfo {year} {2014})},\ \Eprint
  {http://arxiv.org/abs/1402.0233} {arXiv:1402.0233 [hep-ph]} \BibitemShut
  {NoStop}%
%%CITATION = ARXIV:1402.0233;%%
\bibitem [{\citenamefont {Alanne}\ \emph
  {et~al.}(2015{\natexlab{b}})\citenamefont {Alanne}, \citenamefont {Meroni},
  \citenamefont {Sannino},\ and\ \citenamefont {Tuominen}}]{Alanne:2015fqh}%
  \BibitemOpen
  \bibfield  {author} {\bibinfo {author} {\bibfnamefont {T.}~\bibnamefont
  {Alanne}}, \bibinfo {author} {\bibfnamefont {A.}~\bibnamefont {Meroni}},
  \bibinfo {author} {\bibfnamefont {F.}~\bibnamefont {Sannino}}, \ and\
  \bibinfo {author} {\bibfnamefont {K.}~\bibnamefont {Tuominen}},\ }\href@noop
  {} {\  (\bibinfo {year} {2015}{\natexlab{b}})},\ \Eprint
  {http://arxiv.org/abs/1511.01910} {arXiv:1511.01910 [hep-ph]} \BibitemShut
  {NoStop}%
%%CITATION = ARXIV:1511.01910;%%
\bibitem [{\citenamefont {Appelquist}\ \emph {et~al.}(1999)\citenamefont
  {Appelquist}, \citenamefont {Rodrigues~da Silva},\ and\ \citenamefont
  {Sannino}}]{Appelquist:1999dq}%
  \BibitemOpen
  \bibfield  {author} {\bibinfo {author} {\bibfnamefont {T.}~\bibnamefont
  {Appelquist}}, \bibinfo {author} {\bibfnamefont {P.~S.}\ \bibnamefont
  {Rodrigues~da Silva}}, \ and\ \bibinfo {author} {\bibfnamefont
  {F.}~\bibnamefont {Sannino}},\ }\href {\doibase 10.1103/PhysRevD.60.116007}
  {\bibfield  {journal} {\bibinfo  {journal} {Phys. Rev.}\ }\textbf {\bibinfo
  {volume} {D60}},\ \bibinfo {pages} {116007} (\bibinfo {year} {1999})},\
  \Eprint {http://arxiv.org/abs/hep-ph/9906555} {arXiv:hep-ph/9906555 [hep-ph]}
  \BibitemShut {NoStop}%
%%CITATION = HEP-PH/9906555;%%
\bibitem [{\citenamefont {Duan}\ \emph {et~al.}(2001)\citenamefont {Duan},
  \citenamefont {Rodrigues~da Silva},\ and\ \citenamefont
  {Sannino}}]{Duan:2000dy}%
  \BibitemOpen
  \bibfield  {author} {\bibinfo {author} {\bibfnamefont {Z.-y.}\ \bibnamefont
  {Duan}}, \bibinfo {author} {\bibfnamefont {P.~S.}\ \bibnamefont {Rodrigues~da
  Silva}}, \ and\ \bibinfo {author} {\bibfnamefont {F.}~\bibnamefont
  {Sannino}},\ }\href {\doibase 10.1016/S0550-3213(00)00550-2} {\bibfield
  {journal} {\bibinfo  {journal} {Nucl. Phys.}\ }\textbf {\bibinfo {volume}
  {B592}},\ \bibinfo {pages} {371} (\bibinfo {year} {2001})},\ \Eprint
  {http://arxiv.org/abs/hep-ph/0001303} {arXiv:hep-ph/0001303 [hep-ph]}
  \BibitemShut {NoStop}%
%%CITATION = HEP-PH/0001303;%%
\bibitem [{\citenamefont {Ryttov}\ and\ \citenamefont
  {Sannino}(2008)}]{Ryttov:2008xe}%
  \BibitemOpen
  \bibfield  {author} {\bibinfo {author} {\bibfnamefont {T.~A.}\ \bibnamefont
  {Ryttov}}\ and\ \bibinfo {author} {\bibfnamefont {F.}~\bibnamefont
  {Sannino}},\ }\href {\doibase 10.1103/PhysRevD.78.115010} {\bibfield
  {journal} {\bibinfo  {journal} {Phys. Rev.}\ }\textbf {\bibinfo {volume}
  {D78}},\ \bibinfo {pages} {115010} (\bibinfo {year} {2008})},\ \Eprint
  {http://arxiv.org/abs/0809.0713} {arXiv:0809.0713 [hep-ph]} \BibitemShut
  {NoStop}%
%%CITATION = ARXIV:0809.0713;%%
\bibitem [{\citenamefont {Galloway}\ \emph {et~al.}(2010)\citenamefont
  {Galloway}, \citenamefont {Evans}, \citenamefont {Luty},\ and\ \citenamefont
  {Tacchi}}]{Galloway:2010bp}%
  \BibitemOpen
  \bibfield  {author} {\bibinfo {author} {\bibfnamefont {J.}~\bibnamefont
  {Galloway}}, \bibinfo {author} {\bibfnamefont {J.~A.}\ \bibnamefont {Evans}},
  \bibinfo {author} {\bibfnamefont {M.~A.}\ \bibnamefont {Luty}}, \ and\
  \bibinfo {author} {\bibfnamefont {R.~A.}\ \bibnamefont {Tacchi}},\ }\href
  {\doibase 10.1007/JHEP10(2010)086} {\bibfield  {journal} {\bibinfo  {journal}
  {JHEP}\ }\textbf {\bibinfo {volume} {10}},\ \bibinfo {pages} {086} (\bibinfo
  {year} {2010})},\ \Eprint {http://arxiv.org/abs/1001.1361} {arXiv:1001.1361
  [hep-ph]} \BibitemShut {NoStop}%
%%CITATION = ARXIV:1001.1361;%%
\bibitem [{\citenamefont {Minkowski}(1977)}]{Minkowski:1977sc}%
  \BibitemOpen
  \bibfield  {author} {\bibinfo {author} {\bibfnamefont {P.}~\bibnamefont
  {Minkowski}},\ }\href {\doibase 10.1016/0370-2693(77)90435-X} {\bibfield
  {journal} {\bibinfo  {journal} {Phys. Lett.}\ }\textbf {\bibinfo {volume}
  {B67}},\ \bibinfo {pages} {421} (\bibinfo {year} {1977})}\BibitemShut
  {NoStop}%
%%CITATION = PHLTA,B67,421;%%
\bibitem [{\citenamefont {Yanagida}(1979)}]{Yanagida:1979as}%
  \BibitemOpen
  \bibfield  {author} {\bibinfo {author} {\bibfnamefont {T.}~\bibnamefont
  {Yanagida}},\ }\bibfield  {booktitle} {\emph {\bibinfo {booktitle}
  {{Proceedings: Workshop on the Unified Theories and the Baryon Number in the
  Universe, Tsukuba, Japan, 13-14 Feb 1979}}},\ }\href@noop {} {\bibfield
  {journal} {\bibinfo  {journal} {Conf. Proc.}\ }\textbf {\bibinfo {volume}
  {C7902131}},\ \bibinfo {pages} {95} (\bibinfo {year} {1979})},\ \bibinfo
  {note} {[Conf. Proc.C7902131,95(1979)]}\BibitemShut {NoStop}%
%%CITATION = CONFP,C7902131,95;%%
\bibitem [{\citenamefont {Gell-Mann}\ \emph {et~al.}(1979)\citenamefont
  {Gell-Mann}, \citenamefont {Ramond},\ and\ \citenamefont
  {Slansky}}]{GellMann:1980vs}%
  \BibitemOpen
  \bibfield  {author} {\bibinfo {author} {\bibfnamefont {M.}~\bibnamefont
  {Gell-Mann}}, \bibinfo {author} {\bibfnamefont {P.}~\bibnamefont {Ramond}}, \
  and\ \bibinfo {author} {\bibfnamefont {R.}~\bibnamefont {Slansky}},\
  }\bibfield  {booktitle} {\emph {\bibinfo {booktitle} {{Supergravity Workshop
  Stony Brook, New York, September 27-28, 1979}}},\ }\href@noop {} {\bibfield
  {journal} {\bibinfo  {journal} {Conf. Proc.}\ }\textbf {\bibinfo {volume}
  {C790927}},\ \bibinfo {pages} {315} (\bibinfo {year} {1979})},\ \Eprint
  {http://arxiv.org/abs/1306.4669} {arXiv:1306.4669 [hep-th]} \BibitemShut
  {NoStop}%
%%CITATION = ARXIV:1306.4669;%%
\bibitem [{\citenamefont {Mohapatra}\ and\ \citenamefont
  {Senjanovic}(1980)}]{Mohapatra:1979ia}%
  \BibitemOpen
  \bibfield  {author} {\bibinfo {author} {\bibfnamefont {R.~N.}\ \bibnamefont
  {Mohapatra}}\ and\ \bibinfo {author} {\bibfnamefont {G.}~\bibnamefont
  {Senjanovic}},\ }\href {\doibase 10.1103/PhysRevLett.44.912} {\bibfield
  {journal} {\bibinfo  {journal} {Phys. Rev. Lett.}\ }\textbf {\bibinfo
  {volume} {44}},\ \bibinfo {pages} {912} (\bibinfo {year} {1980})}\BibitemShut
  {NoStop}%
%%CITATION = PRLTA,44,912;%%
\bibitem [{\citenamefont {Degrassi}\ \emph {et~al.}(2012)\citenamefont
  {Degrassi}, \citenamefont {Di~Vita}, \citenamefont {Elias-Miro},
  \citenamefont {Espinosa}, \citenamefont {Giudice}, \citenamefont {Isidori},\
  and\ \citenamefont {Strumia}}]{Degrassi:2012ry}%
  \BibitemOpen
  \bibfield  {author} {\bibinfo {author} {\bibfnamefont {G.}~\bibnamefont
  {Degrassi}}, \bibinfo {author} {\bibfnamefont {S.}~\bibnamefont {Di~Vita}},
  \bibinfo {author} {\bibfnamefont {J.}~\bibnamefont {Elias-Miro}}, \bibinfo
  {author} {\bibfnamefont {J.~R.}\ \bibnamefont {Espinosa}}, \bibinfo {author}
  {\bibfnamefont {G.~F.}\ \bibnamefont {Giudice}}, \bibinfo {author}
  {\bibfnamefont {G.}~\bibnamefont {Isidori}}, \ and\ \bibinfo {author}
  {\bibfnamefont {A.}~\bibnamefont {Strumia}},\ }\href {\doibase
  10.1007/JHEP08(2012)098} {\bibfield  {journal} {\bibinfo  {journal} {JHEP}\
  }\textbf {\bibinfo {volume} {08}},\ \bibinfo {pages} {098} (\bibinfo {year}
  {2012})},\ \Eprint {http://arxiv.org/abs/1205.6497} {arXiv:1205.6497
  [hep-ph]} \BibitemShut {NoStop}%
%%CITATION = ARXIV:1205.6497;%%
\bibitem [{\citenamefont {{Peebles}}(1987{\natexlab{a}})}]{Peebles1987}%
  \BibitemOpen
  \bibfield  {author} {\bibinfo {author} {\bibfnamefont {P.}~\bibnamefont
  {{Peebles}}},\ }\href {\doibase 10.1038/327210a0} {\bibfield  {journal}
  {\bibinfo  {journal} {Nature}\ }\textbf {\bibinfo {volume} {327}},\ \bibinfo
  {pages} {210} (\bibinfo {year} {1987}{\natexlab{a}})}\BibitemShut {NoStop}%
\bibitem [{\citenamefont
  {{Peebles}}(1987{\natexlab{b}})}]{1987ApJ...315L..73P}%
  \BibitemOpen
  \bibfield  {author} {\bibinfo {author} {\bibfnamefont {P.}~\bibnamefont
  {{Peebles}}},\ }\href {\doibase 10.1086/184863} {\bibfield  {journal}
  {\bibinfo  {journal} {Astrophys. J. Lett.}\ }\textbf {\bibinfo {volume}
  {315}},\ \bibinfo {pages} {L73} (\bibinfo {year}
  {1987}{\natexlab{b}})}\BibitemShut {NoStop}%
\bibitem [{\citenamefont {Enqvist}\ and\ \citenamefont
  {McDonald}(1999)}]{Enqvist:1998pf}%
  \BibitemOpen
  \bibfield  {author} {\bibinfo {author} {\bibfnamefont {K.}~\bibnamefont
  {Enqvist}}\ and\ \bibinfo {author} {\bibfnamefont {J.}~\bibnamefont
  {McDonald}},\ }\href {\doibase 10.1103/PhysRevLett.83.2510} {\bibfield
  {journal} {\bibinfo  {journal} {Phys. Rev. Lett.}\ }\textbf {\bibinfo
  {volume} {83}},\ \bibinfo {pages} {2510} (\bibinfo {year} {1999})},\ \Eprint
  {http://arxiv.org/abs/hep-ph/9811412} {arXiv:hep-ph/9811412 [hep-ph]}
  \BibitemShut {NoStop}%
%%CITATION = HEP-PH/9811412;%%
\bibitem [{\citenamefont {Enqvist}\ and\ \citenamefont
  {McDonald}(2000)}]{Enqvist:1999hv}%
  \BibitemOpen
  \bibfield  {author} {\bibinfo {author} {\bibfnamefont {K.}~\bibnamefont
  {Enqvist}}\ and\ \bibinfo {author} {\bibfnamefont {J.}~\bibnamefont
  {McDonald}},\ }\href {\doibase 10.1103/PhysRevD.62.043502} {\bibfield
  {journal} {\bibinfo  {journal} {Phys. Rev.}\ }\textbf {\bibinfo {volume}
  {D62}},\ \bibinfo {pages} {043502} (\bibinfo {year} {2000})},\ \Eprint
  {http://arxiv.org/abs/hep-ph/9912478} {arXiv:hep-ph/9912478 [hep-ph]}
  \BibitemShut {NoStop}%
%%CITATION = HEP-PH/9912478;%%
\bibitem [{\citenamefont {Harigaya}\ \emph {et~al.}(2014)\citenamefont
  {Harigaya}, \citenamefont {Kamada}, \citenamefont {Kawasaki}, \citenamefont
  {Mukaida},\ and\ \citenamefont {Yamada}}]{Harigaya:2014tla}%
  \BibitemOpen
  \bibfield  {author} {\bibinfo {author} {\bibfnamefont {K.}~\bibnamefont
  {Harigaya}}, \bibinfo {author} {\bibfnamefont {A.}~\bibnamefont {Kamada}},
  \bibinfo {author} {\bibfnamefont {M.}~\bibnamefont {Kawasaki}}, \bibinfo
  {author} {\bibfnamefont {K.}~\bibnamefont {Mukaida}}, \ and\ \bibinfo
  {author} {\bibfnamefont {M.}~\bibnamefont {Yamada}},\ }\href {\doibase
  10.1103/PhysRevD.90.043510} {\bibfield  {journal} {\bibinfo  {journal} {Phys.
  Rev.}\ }\textbf {\bibinfo {volume} {D90}},\ \bibinfo {pages} {043510}
  (\bibinfo {year} {2014})},\ \Eprint {http://arxiv.org/abs/1404.3138}
  {arXiv:1404.3138 [hep-ph]} \BibitemShut {NoStop}%
%%CITATION = ARXIV:1404.3138;%%
\bibitem [{\citenamefont {Ade}\ \emph {et~al.}(2014)\citenamefont {Ade} \emph
  {et~al.}}]{Ade:2013uln}%
  \BibitemOpen
  \bibfield  {author} {\bibinfo {author} {\bibfnamefont {P.}~\bibnamefont
  {Ade}} \emph {et~al.} (\bibinfo {collaboration} {Planck Collaboration}),\
  }\href {\doibase 10.1051/0004-6361/201321569} {\bibfield  {journal} {\bibinfo
   {journal} {Astron. Astrophys.}\ }\textbf {\bibinfo {volume} {571}},\
  \bibinfo {pages} {A22} (\bibinfo {year} {2014})},\ \Eprint
  {http://arxiv.org/abs/1303.5082} {arXiv:1303.5082 [astro-ph.CO]} \BibitemShut
  {NoStop}%
%%CITATION = ARXIV:1303.5082;%%
\bibitem [{\citenamefont {Ade}\ \emph {et~al.}(2015)\citenamefont {Ade} \emph
  {et~al.}}]{Ade:2015lrj}%
  \BibitemOpen
  \bibfield  {author} {\bibinfo {author} {\bibfnamefont {P.~A.~R.}\
  \bibnamefont {Ade}} \emph {et~al.} (\bibinfo {collaboration} {Planck
  Collaboration}),\ }\href@noop {} {\  (\bibinfo {year} {2015})},\ \Eprint
  {http://arxiv.org/abs/1502.02114} {arXiv:1502.02114 [astro-ph.CO]}
  \BibitemShut {NoStop}%
%%CITATION = ARXIV:1502.02114;%%
\bibitem [{\citenamefont {Weinberg}(2008)}]{Weinberg:2008zzc}%
  \BibitemOpen
  \bibfield  {author} {\bibinfo {author} {\bibfnamefont {S.}~\bibnamefont
  {Weinberg}},\ }\href@noop {} {\emph {\bibinfo {title} {{Cosmology}}}}\
  (\bibinfo  {publisher} {Oxford Univ. Pr.},\ \bibinfo {address} {Oxford, UK},\
  \bibinfo {year} {2008})\ pp.\ \bibinfo {pages} {1--593}\BibitemShut {NoStop}%
%%CITATION = ISBN-9780198526827;%%
\bibitem [{\citenamefont {Kolb}\ and\ \citenamefont
  {Turner}(1990)}]{Kolb:1990vq}%
  \BibitemOpen
  \bibfield  {author} {\bibinfo {author} {\bibfnamefont {E.~W.}\ \bibnamefont
  {Kolb}}\ and\ \bibinfo {author} {\bibfnamefont {M.~S.}\ \bibnamefont
  {Turner}},\ }\href@noop {} {\emph {\bibinfo {title} {{The Early
  Universe}}}},\ \bibinfo {series} {Frontiers in Physics}, Vol.~\bibinfo
  {volume} {69}\ (\bibinfo  {publisher} {Addison-Wesley},\ \bibinfo {address}
  {Reading, MA},\ \bibinfo {year} {1990})\ pp.\ \bibinfo {pages}
  {1--547}\BibitemShut {NoStop}%
%%CITATION = FRPHA,69,1;%%
\bibitem [{\citenamefont {Shaposhnikov}(1987)}]{Shaposhnikov:1987tw}%
  \BibitemOpen
  \bibfield  {author} {\bibinfo {author} {\bibfnamefont {M.~E.}\ \bibnamefont
  {Shaposhnikov}},\ }\href {\doibase 10.1016/0550-3213(87)90127-1} {\bibfield
  {journal} {\bibinfo  {journal} {Nucl. Phys.}\ }\textbf {\bibinfo {volume}
  {B287}},\ \bibinfo {pages} {757} (\bibinfo {year} {1987})}\BibitemShut
  {NoStop}%
%%CITATION = NUPHA,B287,757;%%
\bibitem [{\citenamefont {Shaposhnikov}(1988)}]{Shaposhnikov:1987pf}%
  \BibitemOpen
  \bibfield  {author} {\bibinfo {author} {\bibfnamefont {M.~E.}\ \bibnamefont
  {Shaposhnikov}},\ }\href {\doibase 10.1016/0550-3213(88)90373-2} {\bibfield
  {journal} {\bibinfo  {journal} {Nucl. Phys.}\ }\textbf {\bibinfo {volume}
  {B299}},\ \bibinfo {pages} {797} (\bibinfo {year} {1988})}\BibitemShut
  {NoStop}%
%%CITATION = NUPHA,B299,797;%%
\bibitem [{\citenamefont {Fukugita}\ and\ \citenamefont
  {Yanagida}(1986)}]{Fukugita:1986hr}%
  \BibitemOpen
  \bibfield  {author} {\bibinfo {author} {\bibfnamefont {M.}~\bibnamefont
  {Fukugita}}\ and\ \bibinfo {author} {\bibfnamefont {T.}~\bibnamefont
  {Yanagida}},\ }\href {\doibase 10.1016/0370-2693(86)91126-3} {\bibfield
  {journal} {\bibinfo  {journal} {Phys. Lett.}\ }\textbf {\bibinfo {volume}
  {B174}},\ \bibinfo {pages} {45} (\bibinfo {year} {1986})}\BibitemShut
  {NoStop}%
%%CITATION = PHLTA,B174,45;%%
\bibitem [{\citenamefont {Giudice}\ \emph {et~al.}(2004)\citenamefont
  {Giudice}, \citenamefont {Notari}, \citenamefont {Raidal}, \citenamefont
  {Riotto},\ and\ \citenamefont {Strumia}}]{Giudice:2003jh}%
  \BibitemOpen
  \bibfield  {author} {\bibinfo {author} {\bibfnamefont {G.~F.}\ \bibnamefont
  {Giudice}}, \bibinfo {author} {\bibfnamefont {A.}~\bibnamefont {Notari}},
  \bibinfo {author} {\bibfnamefont {M.}~\bibnamefont {Raidal}}, \bibinfo
  {author} {\bibfnamefont {A.}~\bibnamefont {Riotto}}, \ and\ \bibinfo {author}
  {\bibfnamefont {A.}~\bibnamefont {Strumia}},\ }\href {\doibase
  10.1016/j.nuclphysb.2004.02.019} {\bibfield  {journal} {\bibinfo  {journal}
  {Nucl. Phys.}\ }\textbf {\bibinfo {volume} {B685}},\ \bibinfo {pages} {89}
  (\bibinfo {year} {2004})},\ \Eprint {http://arxiv.org/abs/hep-ph/0310123}
  {arXiv:hep-ph/0310123 [hep-ph]} \BibitemShut {NoStop}%
%%CITATION = HEP-PH/0310123;%%
\bibitem [{\citenamefont {Pilaftsis}\ and\ \citenamefont
  {Underwood}(2004)}]{Pilaftsis:2003gt}%
  \BibitemOpen
  \bibfield  {author} {\bibinfo {author} {\bibfnamefont {A.}~\bibnamefont
  {Pilaftsis}}\ and\ \bibinfo {author} {\bibfnamefont {T.~E.~J.}\ \bibnamefont
  {Underwood}},\ }\href {\doibase 10.1016/j.nuclphysb.2004.05.029} {\bibfield
  {journal} {\bibinfo  {journal} {Nucl. Phys.}\ }\textbf {\bibinfo {volume}
  {B692}},\ \bibinfo {pages} {303} (\bibinfo {year} {2004})},\ \Eprint
  {http://arxiv.org/abs/hep-ph/0309342} {arXiv:hep-ph/0309342 [hep-ph]}
  \BibitemShut {NoStop}%
%%CITATION = HEP-PH/0309342;%%
\bibitem [{\citenamefont {Pilaftsis}\ and\ \citenamefont
  {Underwood}(2005)}]{Pilaftsis:2005rv}%
  \BibitemOpen
  \bibfield  {author} {\bibinfo {author} {\bibfnamefont {A.}~\bibnamefont
  {Pilaftsis}}\ and\ \bibinfo {author} {\bibfnamefont {T.~E.~J.}\ \bibnamefont
  {Underwood}},\ }\href {\doibase 10.1103/PhysRevD.72.113001} {\bibfield
  {journal} {\bibinfo  {journal} {Phys. Rev.}\ }\textbf {\bibinfo {volume}
  {D72}},\ \bibinfo {pages} {113001} (\bibinfo {year} {2005})},\ \Eprint
  {http://arxiv.org/abs/hep-ph/0506107} {arXiv:hep-ph/0506107 [hep-ph]}
  \BibitemShut {NoStop}%
%%CITATION = HEP-PH/0506107;%%
\bibitem [{\citenamefont {Akhmedov}\ \emph {et~al.}(1998)\citenamefont
  {Akhmedov}, \citenamefont {Rubakov},\ and\ \citenamefont
  {Smirnov}}]{Akhmedov:1998qx}%
  \BibitemOpen
  \bibfield  {author} {\bibinfo {author} {\bibfnamefont {E.~K.}\ \bibnamefont
  {Akhmedov}}, \bibinfo {author} {\bibfnamefont {V.~A.}\ \bibnamefont
  {Rubakov}}, \ and\ \bibinfo {author} {\bibfnamefont {A.~{\relax Yu}.}\
  \bibnamefont {Smirnov}},\ }\href {\doibase 10.1103/PhysRevLett.81.1359}
  {\bibfield  {journal} {\bibinfo  {journal} {Phys. Rev. Lett.}\ }\textbf
  {\bibinfo {volume} {81}},\ \bibinfo {pages} {1359} (\bibinfo {year}
  {1998})},\ \Eprint {http://arxiv.org/abs/hep-ph/9803255}
  {arXiv:hep-ph/9803255 [hep-ph]} \BibitemShut {NoStop}%
%%CITATION = HEP-PH/9803255;%%
\bibitem [{\citenamefont {Cannoni}(2014)}]{Cannoni:2013bza}%
  \BibitemOpen
  \bibfield  {author} {\bibinfo {author} {\bibfnamefont {M.}~\bibnamefont
  {Cannoni}},\ }\href {\doibase 10.1103/PhysRevD.89.103533} {\bibfield
  {journal} {\bibinfo  {journal} {Phys. Rev.}\ }\textbf {\bibinfo {volume}
  {D89}},\ \bibinfo {pages} {103533} (\bibinfo {year} {2014})},\ \Eprint
  {http://arxiv.org/abs/1311.4494} {arXiv:1311.4494 [astro-ph.CO]} \BibitemShut
  {NoStop}%
%%CITATION = ARXIV:1311.4494;%%
\bibitem [{\citenamefont {Kusenko}\ \emph {et~al.}(2016)\citenamefont
  {Kusenko}, \citenamefont {Pearce},\ and\ \citenamefont
  {Yang}}]{Kusenko:2016vcq}%
  \BibitemOpen
  \bibfield  {author} {\bibinfo {author} {\bibfnamefont {A.}~\bibnamefont
  {Kusenko}}, \bibinfo {author} {\bibfnamefont {L.}~\bibnamefont {Pearce}}, \
  and\ \bibinfo {author} {\bibfnamefont {L.}~\bibnamefont {Yang}},\ }\href@noop
  {} {\  (\bibinfo {year} {2016})},\ \Eprint {http://arxiv.org/abs/1604.02382}
  {arXiv:1604.02382 [hep-ph]} \BibitemShut {NoStop}%
%%CITATION = ARXIV:1604.02382;%%
\end{thebibliography}%

\end{document}